\documentclass[]{aa} % for a paper on 1 column  
\usepackage{graphicx}
\usepackage{txfonts}
\usepackage{xcolor}
\usepackage{subcaption}         % necessary for continued figures, example in section 3
\usepackage{placeins}           % useful with \FloatBarrier, to keep 
\usepackage[breaklinks, colorlinks, citecolor=blue, linkcolor=blue]{hyperref}
\usepackage{listings}

\definecolor{codegreen}{rgb}{0,0.6,0}
\definecolor{codegray}{rgb}{0.5,0.5,0.5}
\definecolor{codepurple}{rgb}{0.58,0,0.82}
\definecolor{backcolour}{rgb}{0.95,0.95,0.92}

\lstdefinestyle{mystyle}{
    backgroundcolor=\color{backcolour},   
    commentstyle=\color{codegreen},
    keywordstyle=\color{magenta},
    numberstyle=\tiny\color{codegray},
    stringstyle=\color{codepurple},
    basicstyle=\ttfamily\scriptsize,
    breakatwhitespace=false,         
    breaklines=true,                 
    captionpos=b,                    
    keepspaces=true,                 
    showspaces=false,                
    showstringspaces=false,
    showtabs=false,                  
    tabsize=2
}

\newcommand{\norm}[1]{\left\lVert#1\right\rVert}

\newcommand{\code}{\texttt{3DStokesFlow}}
\newcommand{\urlcode}{\url{http://github.com/aasensio/3DStokesFlow}}

\begin{document} 

    \title{\code: simulation-based inference for \\ 3D Stokes profiles using flow matching}
    
    \author{A. Asensio Ramos\inst{1,2}, Kai E. Yang\inst{3}, M. J. Mart\'{\i}nez Gonz\'alez\inst{1,2}, S. Curt Dodds\inst{5}, Xudong Sun\inst{4}}
    
    \institute{
    Instituto de Astrof\'isica de Canarias (IAC), Avda V\'ia L\'actea S/N,
    38200 La Laguna, Tenerife, Spain\\
    \email{andres.asensio@iac.es}
    \and
    Departamento de Astrof\'isica, Universidad de La Laguna, 38205 La Laguna, Tenerife, Spain
    \and
    SETI Institute, Mountain View, CA 94043, USA
    \and
    Institute for Astronomy, University of Hawai‘i at M\=anoa, Pukalani, HI 96768, USA
    \and
    Institute for Astronomy, University of Hawai‘i at M\=anoa, Honolulu, HI 96822, USA
    }

    \date{Received ; accepted }
    
% \abstract{}{}{}{}{} 
% 5 {} token are mandatory
 
    \abstract
    {The standard interpretation of observed Stokes profiles to infer the
    physical conditions of the solar atmosphere is inherently an ill-defined
    problem due to observational noise and mathematical degeneracies.
    Traditional pixel-by-pixel (1D) inversion codes provide point estimates with unreliable
    uncertainties, at the expense of significant computational time.
    Recent machine-learning-based Bayesian frameworks are
    restricted to 1D spatial configurations, ignoring crucial
    spatial correlations between neighboring pixels.}
    {We aim to develop a
    novel multidimensional inversion framework capable of performing fast and scalable
    Bayesian inference across an entire 2D field-of-view (FoV). This approach
    seeks to provide accurate height-dependent atmospheric parameters with
    reliable posterior distributions while exploiting spatial correlations.}
    {We introduce a new generative modeling strategy based on
    conditional flow matching. The model utilizes multi-scale spatial features extracted
    from observed Stokes profiles in the Fe I line pair at 630 nm, which then 
    conditions a flow matching generative model to sample from the 
    complex posterior distribution of the atmospheric parameters. The
    framework is trained using realistic 3D quiet Sun magnetohydrodynamic 
    simulations.}
    {Validation on
    independent synthetic datasets demonstrates that the model accurately
    captures the true 3D stratification of all thermodynamic and magnetic parameters. 
    Because the code additionally provides a geometrical height scale, it allows for the computation
    of 3D electric current density maps, Lorentz forces, and Ohmic and ambipolar
    dissipation maps in the solar photosphere. Application to real Hinode/SP
    quiet Sun observations yields highly localized electric currents at magnetic
    boundaries. We also leverage the 3D geometrical information to trace
    the emergence of small-scale emerging magnetic loops across the solar
    atmosphere.}
    {}
    %{\torchmfbd\ provides a modern, flexible, and efficient framework for MFBD, enabling researchers to push the limits of ground-based astronomical imaging.}
    
    \titlerunning{\code: simulation-based inference for 3D Stokes profiles using flow matching}
    \authorrunning{Asensio Ramos et al.}
    \maketitle
    \nolinenumbers
    
    \abstract

%
%-------------------------------------------------------------------
\section{Introduction}
\label{sec:introduction}
The interpretation of observed Stokes profiles to obtain information about the
physical conditions of the solar atmosphere is routinely performed with
inversion codes \citep{2016LRSP...13....4D,2017SSRv..210..109D}. These codes are
based on the iterative modification of atmospheric models so that the synthetic
Stokes profiles emerging from the model atmosphere match the observed ones.
Different spectral lines require different levels of complexity in the forward
modeling. The simplest codes assume a Milne-Eddington (ME) atmosphere
\citep{1977SoPh...55...47A,2004ASSL..307.....L}, where gradients along the line-of-sight (LOS) are not
relevant. Codes like \texttt{MERLIN} \citep{2007MmSAI..78..148L}, \texttt{MILOS}
\citep{2007A&A...462.1137O}, \texttt{VFISV} \citep{2011SoPh..273..267B}, and
\texttt{pyMilne} \citep{2019A&A...631A.153D} work under this assumption and are,
on average, very fast. When gradients are relevant, the inversion requires the
prescription of a stratification of the physical parameters with height. If the
lines can be safely assumed to form under the local thermodynamic equilibrium
(LTE) approximation, the forward modeling is relatively simple and fast.
Consequently, several codes have been developed to perform inversions under this
assumption. Starting with the seminal work of \cite{1992ApJ...398..375R} and
the development of \texttt{SIR}, a handful of codes are now available, including
\texttt{SPINOR} \citep{2000A&A...358.1109F}, \texttt{FIRTEZ}
\citep{2019A&A...629A..24P}, and \texttt{SPINOR-2D} \citep{2012A&A...548A...5V}.
More complex non-LTE line formation is required for the inversion of
chromospheric lines. These approaches need to consistently solve the radiative
transfer equation and the statistical equilibrium equations at each iteration of
the inversion process. These codes are much more complex and computationally
demanding, and only a few are available to the community: \texttt{NICOLE}
\citep{2015A&A...577A...7S}, \texttt{STiC} \citep{2019A&A...623A..74D},
\texttt{SNAPI} \citep{2018A&A...617A..24M}, \texttt{DeSIRe}
\citep{2022A&A...660A..37R}, and \texttt{TIC} \citep{2022ApJ...933..145L}.

Despite all the efforts carried out in recent decades, it is important to note
that the inversion of Stokes profiles remains an ill-defined problem. This is
not just a consequence of potential mathematical degeneracies in the forward
modeling (such as the well-known 180° ambiguity in the azimuth of the magnetic
field), in which identical Stokes profiles can be produced by different
atmospheric models. The presence of noise in observations also induces
non-unique solutions, producing uncertainties in the inferred atmospheric
parameters. Standard inversion codes often provide a point estimate of the final
covariance matrix at the solution \citep[e.g., see][]{1992ApJ...398..375R}, but
the inferred uncertainties obtained from this covariance matrix tend to be
unreliable because they are calculated purely based on the goodness of the fit
and the parameter sensitivity of the problem. For this reason, strictly
speaking, the inversion of Stokes profiles must be treated as a probabilistic
inference problem. The goal is to characterize the posterior distribution of the
atmospheric parameters given the observed Stokes profiles. This was first
attempted by \cite{2007A&A...476..959A} using a Markov Chain Monte Carlo (MCMC)
method under the assumption of a ME atmosphere. Given that this approach is
computationally demanding, it has been difficult to extend it to more complex
atmospheres.

Only recently, with the advent of machine learning techniques, has it been
possible to perform Bayesian inference in more complex atmospheres. The first
attempt was performed by \cite{2019ApJ...873..128O}, who used invertible neural
networks \citep{2018arXiv180804730A} to capture degeneracies and uncertainties
when inverting Stokes profiles from flaring regions. More recently,
\cite{2022A&A...659A.165D} used normalizing flows to provide extremely fast
variational approximations to the posterior distribution in lines formed in LTE
and non-LTE. \cite{2023SoPh..298...98M} trained a neural model to provide a
Gaussian variational approximation to the posterior distribution, while
\cite{2024ApJ...977..101X} trained emulators to accelerate the forward modeling
and then used MCMC methods to sample the posterior distribution. Finally, the
latest effort in this direction was presented by \cite{2025A&A...703A..55A}, who
treated inversions as a neural translation problem and trained a model to
directly map Stokes profiles to atmospheric parameters in a generative fashion, thereby providing a
variational approximation to the posterior distribution.

All the variational approximations mentioned above are currently limited to 1D
model atmospheres. However, it is becoming increasingly clear that exploiting
the spatial correlation between neighboring pixels is crucial to obtaining more
accurate and reliable inferences
\citep{2019A&A...626A.102A,2024ApJ...976..204Y}. Given the dimensionality of the
resulting posterior distribution when inverting 2D FoVs,
performing Bayesian inference in this case is an extremely challenging problem.
The dimensionality of the problem grows with the number of pixels multiplied by
the depth stratification of all physical parameters. For instance, a 2D map
of 100$\times$100 pixels with a depth stratification of 10 points for temperature, 
LOS velocity, and the three cartesian components of the magnetic field, results in 
a posterior distribution of dimension $5 \times 10 \times 100 \times 100$, 
i.e., half a million dimensions.

In this work, we take full
advantage of recent advances in generative modeling to propose a novel strategy
based on flow matching to perform Bayesian inference of atmospheric parameters
from Stokes profiles. This method exploits the spatial
correlation between neighboring pixels and can provide accurate inferences and
uncertainties for the atmospheric parameters across the entire observed FoV.
Apart from a significant improvement in terms of speed, the exploitation of
spatial correlation allows us to obtain information regarding geometrical
height and gas pressure. This opens up the possibility of using very simple disambiguation 
methods for the transverse components of the magnetic field, which allows us to calculate
electric current density maps. From this information, we demonstrate that it is
possible to obtain maps of the ohmic and ambipolar dissipation in the
photosphere purely from observations\footnote{\url{https://aasensio.github.io/3dstokesflow}}.

\section{Bayesian Inference via Flow Matching}
\label{sec:inverse_problem}

\subsection{Bayesian Inference for Stokes profiles inversion}
The inference of the atmospheric parameters from a 2D FoV of observed Stokes profiles can be framed as a Bayesian inference problem.
Let us denote by $P \in \mathbb{R}^{N \times d}$ the high-dimensional vector that contains the atmospheric parameters 
(of dimension $d$) for each of the $N$ pixels across the entire FoV. We observe the Stokes profiles of some selected
spectral lines and denote them by $S \in \mathbb{R}^{N \times m}$, where $m$ denotes the product of all 
sampled wavelength points across the four Stokes parameters. The goal is to characterize the posterior 
distribution $p(P|S)$, which captures all the information about the atmospheric parameters given the observed data.
The Bayes theorem allows us to express this posterior distribution as:
\begin{equation}
p(P|S) = \frac{p(S|P) p(P)}{p(S)},
\end{equation}
where $p(S|P)$ is the likelihood function that captures the forward modeling of the 
Stokes profiles given the atmospheric parameters, $p(P)$ is the prior distribution that encodes our 
prior knowledge about the atmospheric parameters, and $p(S)$ is the evidence that serves as a normalization constant.

From a generative modeling perspective, one can solve the Bayesian inference problem by drawing 
samples from the complex, unknown posterior distribution $p(P|S)$. However, having 
access to the posterior distribution is, in general, intractable. On one hand, the likelihood function $p(S|P)$ is 
complex and computationally expensive to evaluate, especially when the forward modeling involves solving the 
radiative transfer equation under non-LTE conditions. On the other hand, the prior distribution $p(P)$ (models compatible with those
one could find in the Sun) can also be difficult to specify accurately, given the high dimensionality of the parameter 
space and the potential correlations between parameters. 
However, if one has access to a finite dataset of pairs of atmospheric parameters and their corresponding Stokes profiles, 
$\{(P_i, S_i)\}_{i=1}^N$, drawn from the joint distribution $p(P, S)$,
one can leverage this dataset to learn a generative model that approximates the posterior distribution.
Such dataset can simply be generated using simulations, i.e., by running a forward model 
(e.g., an LTE or non-LTE radiative transfer code) on a set of atmospheric models that are representative of the solar atmosphere.

\subsection{Flow Matching}
Flow models \citep{lipman24} are generative models that learn a continuous transformation (or flow) that maps a simple base multivariate
distribution (e.g., a standard Gaussian $\mathcal{N}(0, I)$, with $I$ being the identity matrix) to a complex target distribution.
Let us denote by $x \in \mathbb{R}^M$ a random variable that follows the base distribution.
One can define a time-dependent probability density path $p_t$ for an auxiliary time $t \in [0, 1]$, where 
$p_0$ (at $t=0$) is the tractable base distribution and $p_1$ (for $t=1$) is the target posterior distribution. 
The flow $\psi_t(x)$ is a time-dependent mapping that transports samples $X_0 \sim p_0$ from the base 
distribution to the target distribution, so that $X_1 \sim p_1$.
In general, the trajectory followed by a sample $X_0$ under the flow can be expressed as $X_t = \psi_t(X_0)$. 
The distribution of all samples generated by the flow at any arbitrary time $t$ are, therefore, distributed 
according to the distribution $p_t$ (i.e., $X_t \sim p_t$). 

Instead of directly learning the flow $\psi_t(x)$, it is more common to learn the 
underlying vector field $u_t(x)$ that generates the flow, according to the following ordinary differential equation:
\begin{equation}
\frac{d}{dt} \psi_t(x) = u_t(x), \qquad \psi_0(x) = x.
\label{eq:flow_ode}
\end{equation}
The goal of flow matching is to learn a parameterized function $v_t^\theta(x)$, typically given
as a sufficiently capable neural network, that generates
a flow that produces a probability path $p_t$ from $p_0$ to $p_1$. 
The ideal objective would be to directly minimize the expected mean squared error between our neural vector field and the true target vector field:
\begin{equation}
\mathcal{L}_{FM}(\theta) = \mathbb{E}_{t,X_t} \norm{v_t^\theta(X_t) - u_t(X_t)}^2, 
\quad t \sim \pi(t), X_t \sim p_t
\end{equation}
where $u_t(x)$ is the true target vector field that generates the probability path $p_t$. The 
expectation is taken over the time $t$, which is sampled from the interval $[0, 1]$ with
probability $\pi(t)$, and over the 
samples $X_t$ drawn from the distribution $p_t$. The main problem of this approach is that
we do not have direct access to the true target 
vector field $u_t(x)$ because it is, indeed, dependent on the intractable transformation between
the base distribution and the complex posterior distribution.

To bypass this intractability, \cite{lipman24} used a strategy based on 
analyzing what happens if the target distribution is 
given by a Dirac delta distribution centered 
at a single data point $X_1$ (i.e., $p_1(x) = \delta_{x_1}(x)$). In other words, they 
analyzed the case conditional on a single data point $X_1$. Of all the possible paths that can transport the base 
distribution to this Dirac delta target distribution, we choose the one that is given by the 
optimal transport (OT) between the two distributions. This allows us to define the simple and linear path such 
that $X_t = (1-t)X_0 + t X_1$, where $X_0 \sim p_0$ is a sample drawn from the base distribution (from a
practical point of view, we use samples from the standard normal distribution, so that $p_0=\mathcal{N}(0,I)$). 
According to Eq. (\ref{eq:flow_ode}), the vector field that generates this path 
can be easily calculated as:
\begin{equation}
u_t(x|X_1) = X_1 - X_0.
\label{eq:conditional_vector_field}
\end{equation}
This vector field is constant in time and points directly from the sample $X_0$ to the 
target sample $X_1$. Once trained, the flow generated by this vector field is a straight line between the point $X_0$ 
and the data point $X_1$, also known as the optimal transport path between the two distributions. 
The full loss function for this conditional flow matching strategy can be obtained by averaging over
the data distribution $p_1$ and all intermediate points $X_t$ along the path, resulting in:
\begin{align}
    \mathcal{L}_{CFM}(\theta) &= \mathbb{E}_{t,X_1,X_t} \norm{v_t^\theta(X_t) - u_t(X_t|X_1)}^2, \nonumber \\
    t &\sim \pi(t), X_1 \sim p_1, X_0 \sim p_0, X_t \sim p_t.
\end{align}
Using Eq. (\ref{eq:conditional_vector_field}), the conditional loss can be finally rewritten as:
\begin{align}
    \mathcal{L}_{CFM}(\theta) &= \mathbb{E}_{t,X_1,X_0} \norm{v_t^\theta(X_t) - (X_1-X_0)}^2, \nonumber \\
    t &\sim \pi(t), X_1 \sim p_1, X_0 \sim p_0.
    \label{eq:conditional_loss}
\end{align}

\cite{lipman24} showed that this construction via conditional flow matching gives a 
deceptively simple loss function which is equivalent to minimizing the original flow 
matching loss $\mathcal{L}_{FM}(\theta)$. Consequently, the model $v_t^\theta(x)$, trained
by optimizing Eq. (\ref{eq:conditional_loss}), approximates 
the true target vector field $u_t(x)$, without ever having direct access to it.
Training then proceeds by sampling a data point $X_1$ from the target distribution, sampling a 
noise point $X_0$ from the base distribution, and sampling a time step $t$ from the $\pi(t)$ 
distribution\footnote{In this work, we use a uniform distribution over $[0, 1]$, so that $\pi(t) = \mathcal{U}(0,1)$}. 
The intermediate point $X_t$ is then calculated as $X_t = (1-t)X_0 + t X_1$, and the loss is computed against 
the known vector field $X_1 - X_0$.

The selection of the pair $(X_0, X_1)$ is often relevant for ending up with an efficient model. The
selection is done via a user-defined joint distribution $\pi(X_0, X_1)$, known as the coupling.
The simplest case is to select $X_0$ and $X_1$ independently from their respective distributions, so that the
coupling factorizes as $\pi(X_0, X_1) = \pi(X_0) \pi(X_1)$. 
In this case, it is likely that the path between $X_0$ and $X_1$ will intersect with 
the paths of other data points, which can make the learning task more difficult for the neural network.
Although solutions to end up with non-intersecting paths have been proposed in the literature (such as using a coupling
that enforces optimal transport), we have found great results simply picking uniformly random samples from $p_0$ and $p_1$.
In other words, we choose $X_0 \sim N(0,I)$ and $X_1$ at random from the training set.
%, that we use in this work. Minibatch-OT consists of selecting a 
%batch of data points and a batch of noise points, and then
%solving an optimal transport problem to find the best matching between the two batches, so that the paths
%between the matched pairs do not intersect. (CHECK!!!!!!)

\subsection{External Conditioning}
The velocity model $v_t^\theta(x)$ can be easily extended to include additional conditioning information. 
In the context of Bayesian inference for Stokes profiles inversion, 
one can train a flow matching model to compute the posterior distribution of the atmospheric parameters 
conditioned on the observed Stokes profiles by optimizing the following loss function:
\begin{align}
    \mathcal{L}_{CFM}(\theta) &= \mathbb{E}_{t,X_0,(P,S)} \norm{v_t^\theta(X_t, S) - (P-X_0)}^2, \nonumber \\
    t &\sim \pi(t), (P, S) \sim p(P,S), X_0 \sim p_0,
\end{align}
where the pair $(P, S)$ is drawn from the joint distribution of atmospheric parameters and Stokes 
profiles.

\subsection{Inference}
When the model is trained, inference can be performed by sampling from the base distribution and
then integrating the learned vector field conditioned on the context $S$ to transport the sample 
from the base distribution to the target distribution:
\begin{equation}
P = X_0 + \int_{0}^{1} v_t^\theta(X_t, S) dt.
\label{eq:inference}
\end{equation}
This can be achieved by using any off-the-shelf ODE solver and, given that the probability paths
are smooth, it can be done with a small number of function evaluations, making the inference process very efficient.

\begin{figure*}
    \centering
    \includegraphics[width=\textwidth]{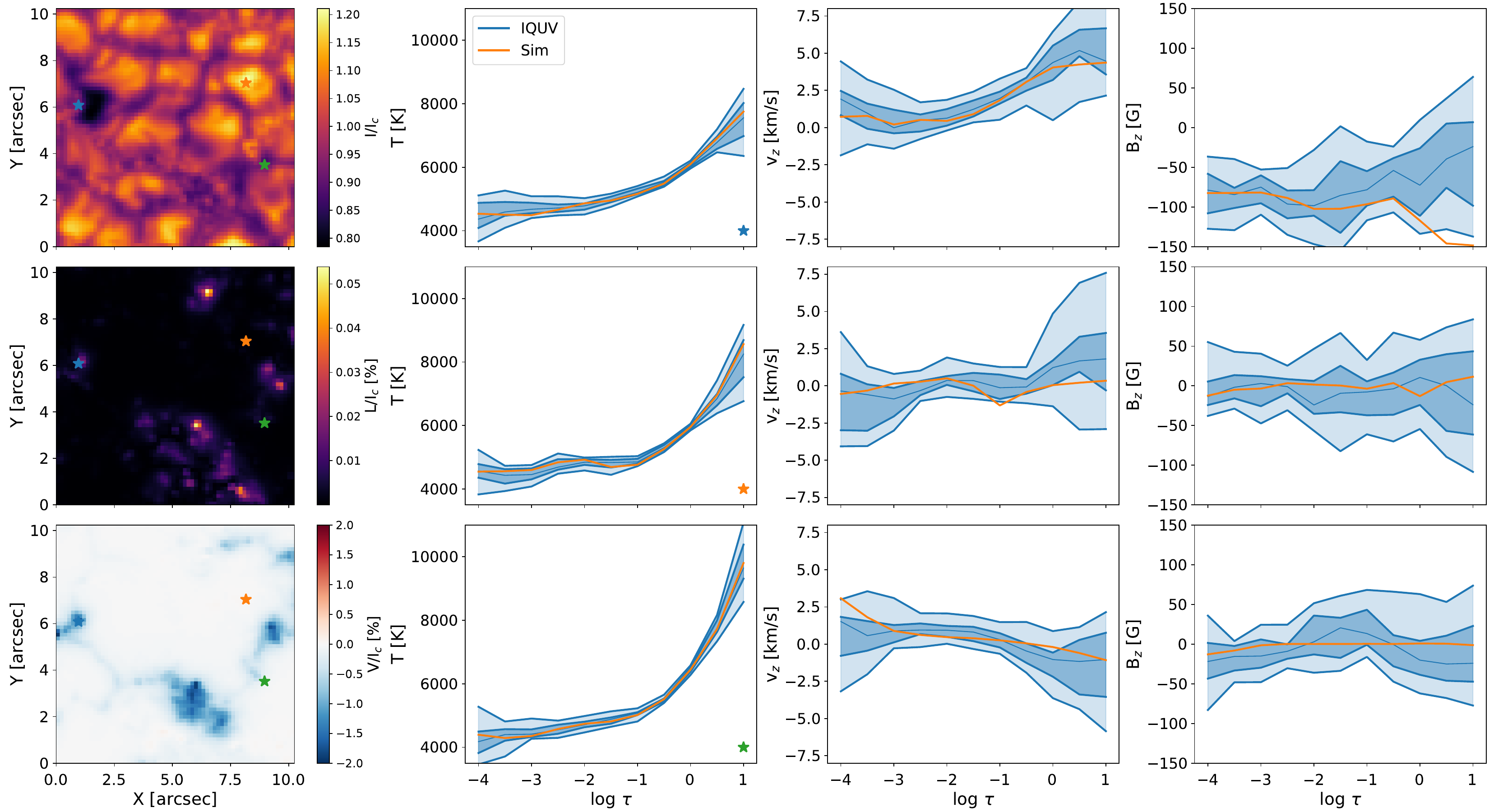}
    \caption{Vertical cuts of the inferred atmospheric parameters 
    (temperature, LOS velocity, and vertical magnetic field) for three columns of the validation 
    snapshot (each row shows a different pixel). The true values from the simulation are shown in orange, while the median 
    of the posterior distribution is shown in thin blue curves. The shaded areas
    correspond to the intervals between the 25th and 75th percentile (dark blue) and the
    5th and 95th percentiles (light blue) of the posterior distribution.\label{fig:vertical_validation} }
\end{figure*}

\section{Training}
\subsection{Training data}
We train the model with the quiet Sun simulations presented in the SPIn4D\footnote{\url{https://ifauh.github.io/SPIN4D}} 
project \citep{Yang_2024,2025ApJ...995..146Y}. These simulations 
were performed with the MURaM code \citep{2005A&A...429..335V,2009ApJ...691..640R,2012ApJ...750...62R,2014ApJ...789..132R}, taking
the quiet Sun simulation of \cite{2014ApJ...789..132R} as the initial condition and evolving them
after adding different magnetic fields to produce snapshots from very quiet Sun to strong plages. The six initial
magnetic configurations considered in this work are: i) zero magnetic field, ii) vertical field with 50 G, iii) inclined 
fields with 50 G in each cartesian component, iv) vertical field with 100 G, v) vertical field with 200 G, vi) vertical 
fields of 200 G, $-150$ G and $-50$ G added to three quadrants of the simulation box to mimic mix polarities see 
\citep[see][for more details on the simulations]{Yang_2024}. The first five configurations are used for
training, while the sixth one with mixed polarities is used for validation and testing. The horizontal simulation pixel
is 16 km and spans $24.6\times 24.6$ Mm horizontally. Snapshots every 12 min are considered, resulting in 
roughly 30 snapshots per simulation. A second set of snaphots is obtained by flipping the sign of the magnetic
field vector, to avoid potential biases in the training data.

The snapshots contain the vertical stratification of the atmospheric parameters as a function of position
on the 2D grid. We consider 
the temperature $T$, LOS velocity $v$, the gas pressure $P$, and the three cartesian components of the magnetic 
field $\mathbf{B}=(B_x,B_y,B_z)$.
The 180$^\circ$ ambiguity in the transverse components of the magnetic field is avoided \citep{2019A&A...626A.102A} 
by transforming the $B_x$ and $B_y$
components of the magnetic field to $B_{p1}=B_t \cos 2 \phi$ and $B_{p2}=B_t \sin 2 \phi$, where $B_t^2=B_x^2+B_y^2=B_{p1}^2+B_{p2}^2$
is the square of the transverse component of the magnetic field, and $\phi=\arctan(B_y/B_x)$ is the azimuth. We found this
solution to be very stable and provide good results.

The original simulation pixel of 16 km is rebinned to the Hinode spatial sampling of 0.16" per pixel, resulting
in a pixel size of 116 km. No previous convolution with the spatial PSF of the telescope is applied to the physical 
parameters, so that the output of the trained model will be deconvolved (unaffected by the PSF) atmospheric parameters, in fashion 
similar to \cite{2019A&A...626A.102A}. Horizontal cuts at constant
$\log \tau_{500}$ between -4 and 1 in steps of 0.5 are extracted. This results in 11 depth points for each 
of the physical parameters, so that the total dimension of the parameter space per column is $d=77$.

The Stokes profiles in the Fe \textsc{i} 6301.5 and 6302.5 \AA\ lines are synthesized at disk center\footnote{This is a current
limitation of the trained model, which can only analyze observations close to disk center. This can be trivially solved by
using other heliocentric angles in the training set.} under the LTE approximation using a 
parallel version of SIR \citep{1992ApJ...398..375R,2019A&A...626A.102A}. We convolve the emergent Stokes profiles with 
the measured spatial and spectral point spread functions (PSF) of Hinode/SOT \citep{2008SoPh..249..167T,2008A&A...484L..17D}. The
results are resampled to the spectral (21.5 m\AA\ per pixel) and spatial sampling (0.16" per pixel) of the Hinode/SP instrument. 
This results in 112 wavelength points for each Stokes parameters. Gaussian noise with a standard deviation of $10^{-3}$ is added 
to the profiles, which is a typical value for Hinode/SP observations.
We finally point out that the line formation region of the Fe \textsc{i} lines almost never reaches $\log \tau_{500}=-4$. However
we included this depth point in the training set and we caution that the information obtained at these
heights is purely learned from the statistical correlations present in the training data conditioned
on the observed Stokes profiles.

\begin{figure*}
    \centering
    \includegraphics[width=\textwidth]{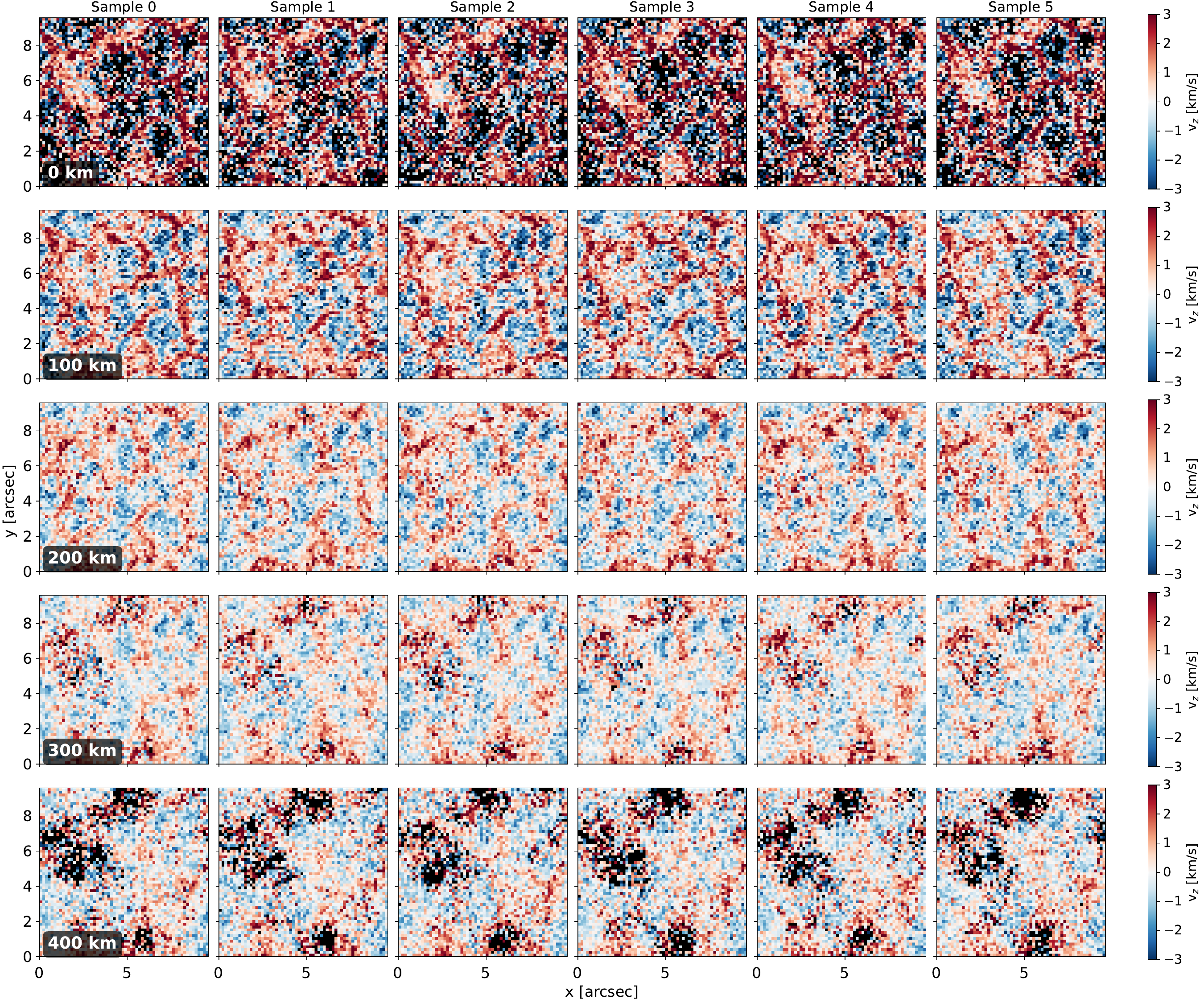}
    \caption{Five samples from the posterior distribution for the vertical velocity computed with the flow
    matching generative model conditioned on the observed Stokes profiles.\label{fig:samples_validation} }
\end{figure*}

\subsection{Architecture and training details}
A total of $10^5$ training patches of size 64$\times$64 pixels are extracted from the original snapshots, randomly
choosing among the snapshots considered for training. The location of the patches is randomly selected, 
ensuring that they are fully contained within the original simulation box. Random horizontal and vertical
flips, together with random rotations in multiples of 90$^\circ$ are applied to the patches to augment the 
training data. Each patch is then a tensor $C \times H \times W$, where $C=77$ is the number of channels
corresponding to the physical parameters, and $H=W=64$ are the height and width of the patch in pixels.
The corresponding Stokes profiles are also extracted in patches of the same size, resulting in a tensor of
shape $L \times H \times W$, where $L=448$ is the concatenation of the four Stokes 
parameters (112 wavelength points each).

The median and standard deviation of the physical parameters are computed for the training dataset and stored. During
training, the physical parameters are normalized by subtracting the median and dividing by three times the standard 
deviation. The Stokes profiles are also normalized. Stokes $I$ is normalized by dividing by 0.5 and subtracting 0.5. 
Stokes $Q$ and $U$ are normalized by multiplying by 100, while Stokes $V$ is normalized by multiplying by 10. We have
found this normalization to be stable during training, giving good results. We note that the normalization of
the Stokes parameters $Q$, $U$ and $V$ is always problematic, given their large dynamic range.

\begin{figure*}
    \centering
    \includegraphics[width=0.48\textwidth]{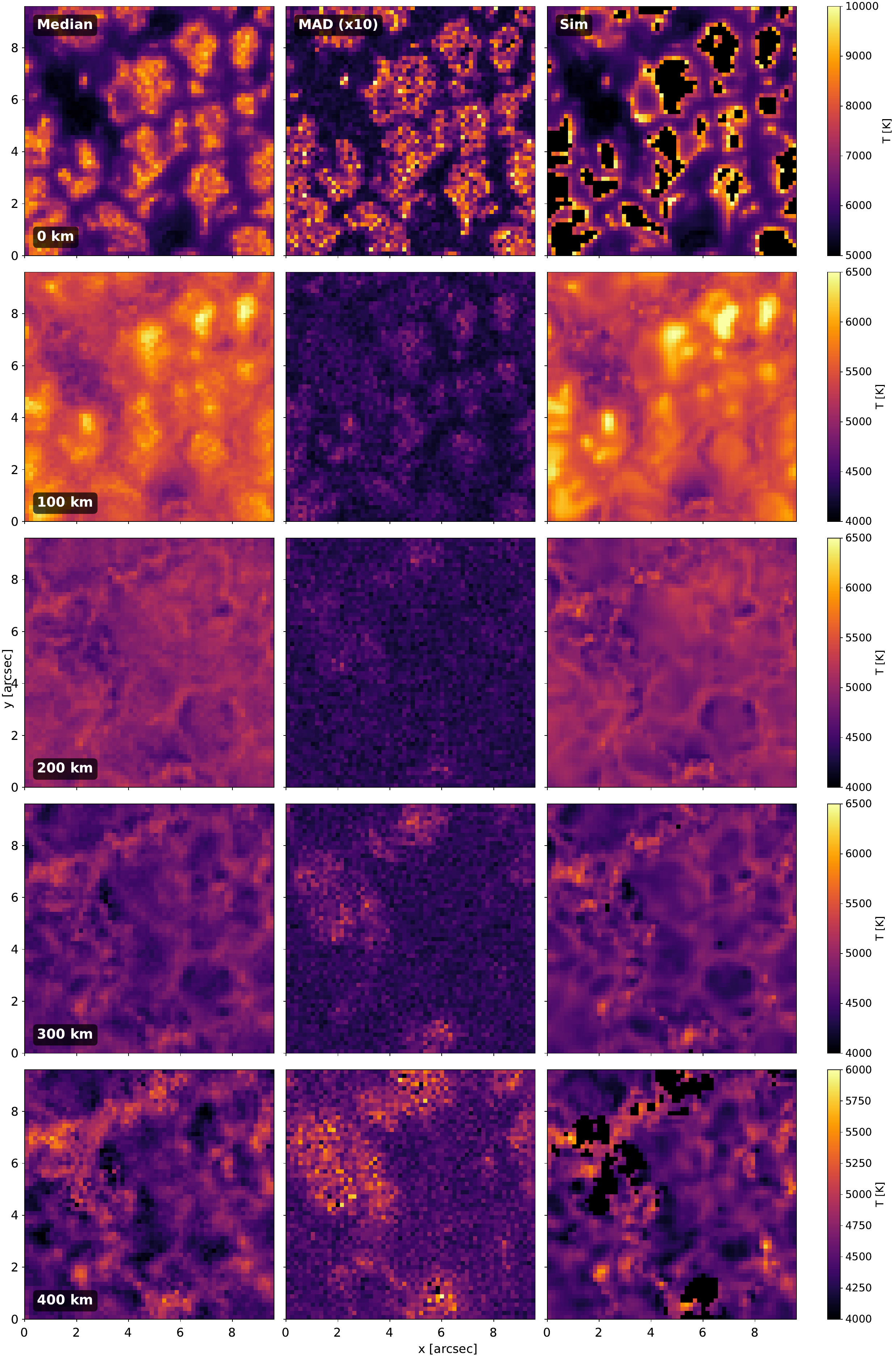}%
    \includegraphics[width=0.48\textwidth]{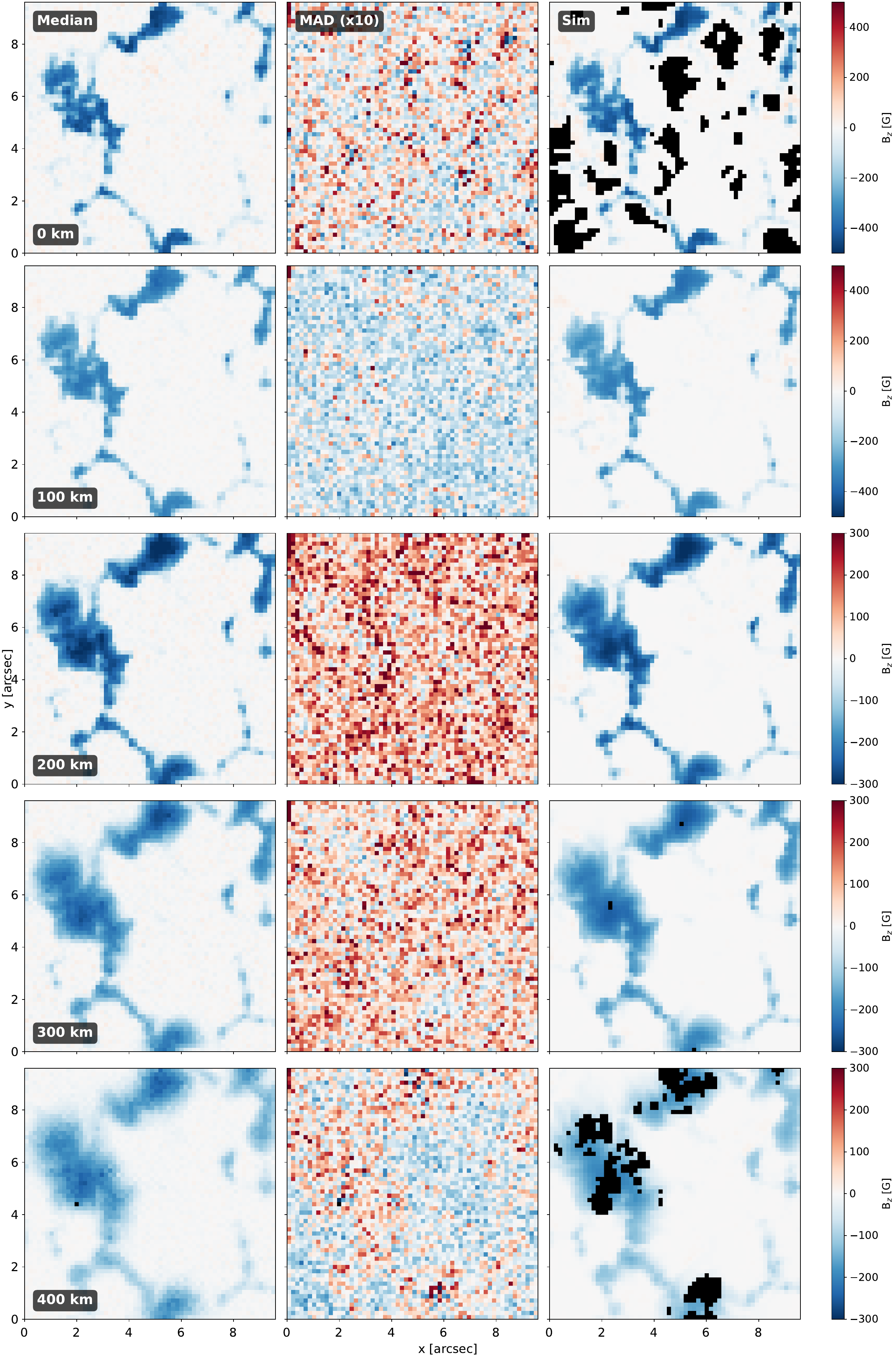}
    \caption{Inferred height-dependent temperature (left three panels) and vertical magnetic field (right three panels). 
    The true values from the simulation are shown in the third column of each panel, while the median and
    MAD of the output probability distributions are shown in the first and second columns of each panel.
    Note that the MAD is multiplied by a factor of 10 so that all panels share the same color bar.\label{fig:validation_maps}}
\end{figure*}

The Stokes profiles are then processed by a conditioning U-Net \citep{unet15} to extract a latent representation that
summarizes the information contained in the Stokes profiles. We choose to use a U-Net architecture for the 
conditioning model because of its ability to 
capture multi-scale spatial features, which are relevant for the prediction of the atmospheric parameters.
This U-Net takes as input the Stokes profiles as a tensor of shape $L \times H \times W$
and produces a latent representation of shape $Z \times H \times W$. We empirically choose $Z=64$
as the number of channels in the latent representation, similar to $C=77$, with the idea of preserving enough
information for contidioning for the velocity model. However, this is a hyperparameter that can be surely optimized.

\begin{figure*}
    \centering
    \includegraphics[width=\textwidth]{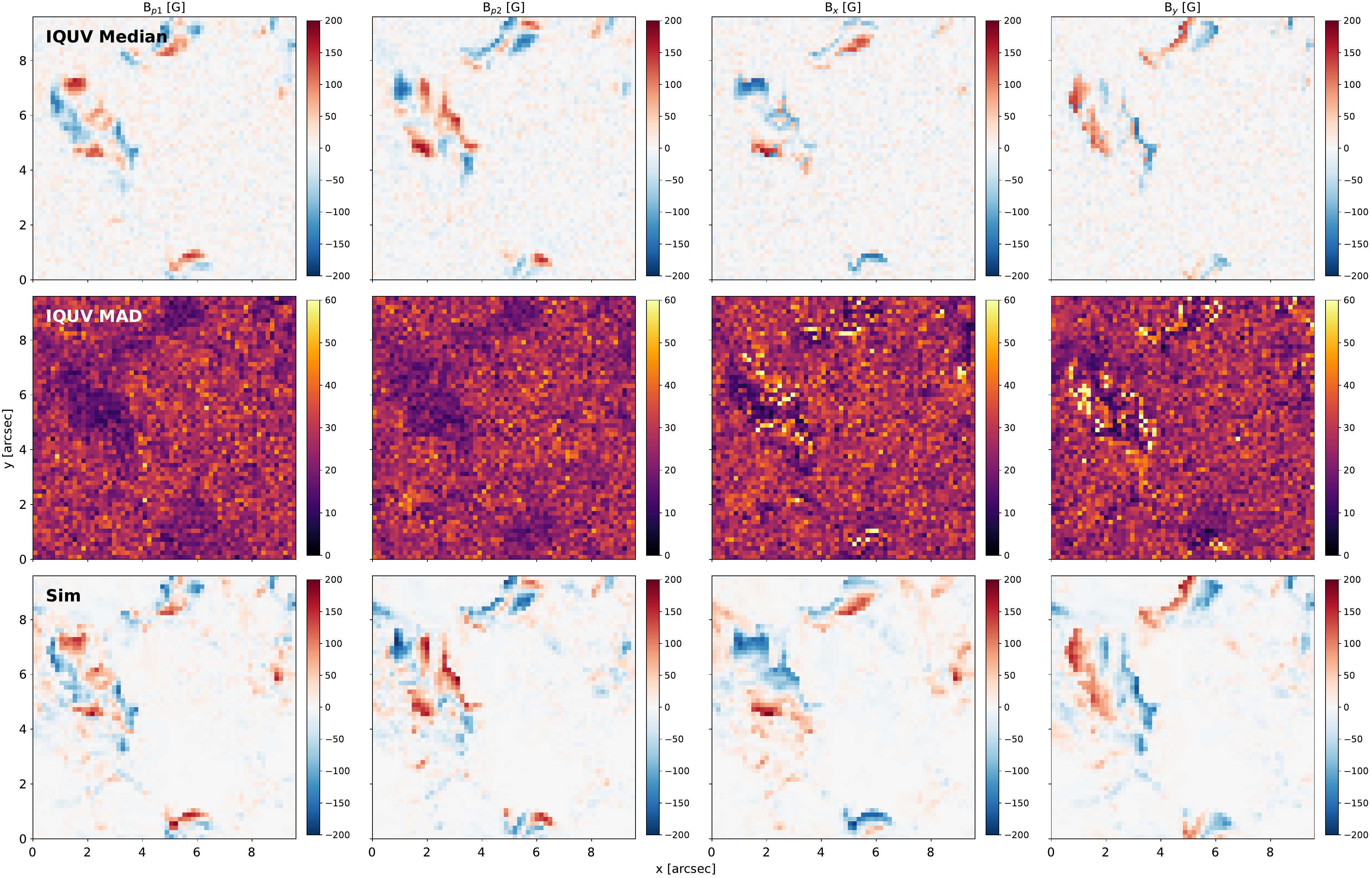}
    \caption{\label{fig:validation_Bxyz} Inferred ambiguous (left two panels) and disambiguated (right two panels) transverse 
    components of the magnetic field for 100 km above the height at which the average optical depth is unity. The median of the 
    posterior is shown in the first row, the MAD of the posterior is
    shown in the second row, and the true values from the simulation are shown in the third row.\label{fig:validation_Bxyz}}
\end{figure*}

The velocity model $v_t^\theta(X_t, S)$ is another U-Net which takes as input 
a tensor of size $D \times H \times W$, where $D=77+64$ is the number of channels corresponding to the 
concatenation of the current state $X_t$ and the learned representation of the conditioning Stokes profiles $S$. 
The time step $t$ is also embedded using a sinusoidal positional embedding and concatenated with the input of 
the velocity U-Net.

The conditioning and velocity models are implemented with the \texttt{PyTorch} package, and trained on a single 
NVIDIA RTX 4090 GPU using half precision numbers. We train for 20 epochs with a batch size of 64 using the 
Adam optimizer \citep{adam14} with an initial learning rate of $3 \times 10^{-4}$, which is slowly annealed during 
training using a cosine law until reaching $9 \times 10^{-5}$. The time per epoch is around 5 min for
a total training time of around 1.5 hours, which is very fast for a model of this complexity and dimensionality.

\section{Validation}
\subsection{Maps of atmospheric parameters}
Once trained, the model is applied to the validation set, the snapshot with mixed polarities 
that was not used during training. The ODE of Eq. (\ref{eq:inference}) is solved with the learned velocity model 
to transport 25 samples from the base distribution to the target distribution. This produces 25 samples of the
3D stratification of the atmospheric parameters, evaluated at the fixed optical depth points defined in 
the training data. The posterior distribution of the atmospheric parameters can be characterized by analyzing 
the statistics of these samples. Figure \ref{fig:vertical_validation} shows the vertical stratification with optical depth
of three different columns in the FoV, marked with color symbols in the first panel. The first column shows the continuum 
intensity normalized to the average continuum, together with the total linear polarization and circular polarization in
the wing of the line. The thin blue curves in the rest of panels display the median (50th percentile) of the posterior
distribution for each optical depth and physical quantity, while the shaded regions corresponds to the area 
between the 25th and 75th (dark blue) and the 5th and 95th percentiles (light blue). The true values from the 
simulation are shown in orange. The results show 
that the model is able to capture the true values within the considered percentiles of the posterior
distribution. Larger uncertainties are found in the lower and upper layers of the atmosphere, which is expected 
given that the lines are mostly sensitive to the region between $\log \tau_{500}=-2.5$ and $\log \tau_{500}=0$. 
The code reliably produces a median value for $B_z$ close to the true value for the most magnetized
pixel (blue star), and returns something compatible with zero for the less magnetized regions (orange and green stars).

\begin{figure*}
    \centering
    \includegraphics[width=\textwidth]{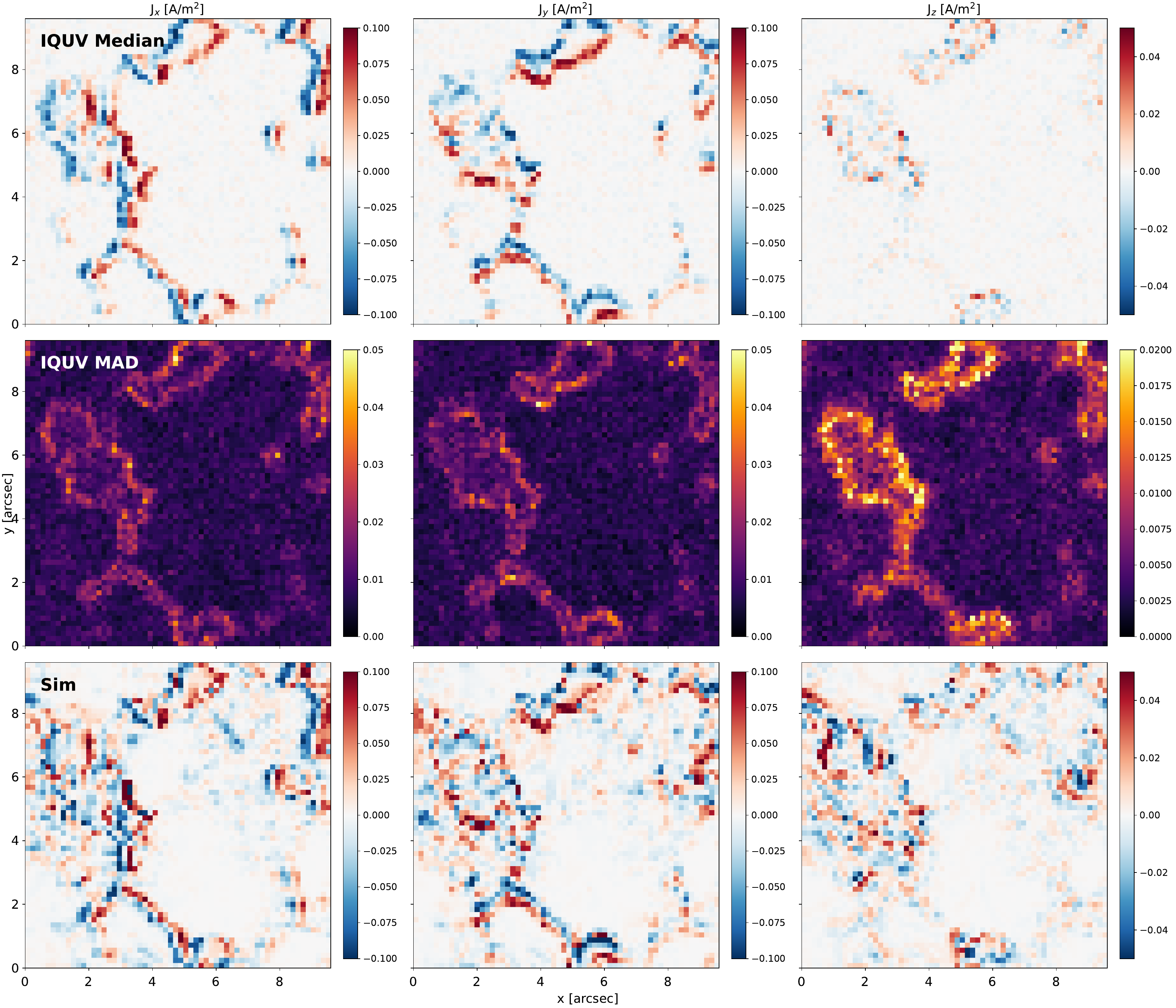}
    \caption{Divergence of the magnetic field (left) and the three cartesian components of the electric current 
    density (right three panels) obtained after disambiguating the transverse components of the magnetic field
    at 100 km. The median and MAD of the posterior are shown in the first and second rows, respectively. The
    last rows shows the inference in the MHD snapshot.\label{fig:validation_currents}}
\end{figure*}

Given that the model produces an estimation of the height of each optical depth surface, it is possible
to use interpolation to obtain the inferred atmospheric parameters as a function of geometrical height
\citep[see, e.g.,][]{2019A&A...626A.102A}. Figure \ref{fig:samples_validation} shows five different samples from 
the posterior distribution for the vertical velocity at five different geometrical heights. 
The changes in optical depth naturally occuring 
due to changes in the temperature and density, together with the presence of Wilson depressions
in the presence of magnetic fields, produce that each column in the FoV has a different mapping
between optical depth and geometrical height. This induces that the physical properties of granules
very deep in the atmosphere are missing (marked as black regions in the panels at $0$ km), while 
the physical properties of intergranular lanes and magnetic elements are missing in the upper geometrical 
layers of the atmosphere (marked as black regions in the panels at 400 km).
There is a significant amount of variability among the samples, which is a 
consequence of the Bayesian nature of the simulation-based model.
Those regions for which the model uncertainty is smaller vary less between samples.
Note that the mapping between optical depth and geometrical height is also a random variable sampled from the
posterior and, as such, different for each sample. Therefore, the interpolation needs to be carried out independently 
for each sample.

Robust statistical summaries of the posterior distribution can be obtained
directly from the samples. Figure \ref{fig:validation_maps} shows the inferred
median and median absolute deviation (MAD) of the posterior distribution for
temperatures and vertical magnetic fields as a function of geometrical height,
alongside their true values in the validation snapshot. Because we crop the
simulations at $\log \tau_{500}=1$ and $\log \tau_{500}=-4$, we lack access to
the temperature and velocity in certain parts of the Field of View (FoV) at 0
and 400 km for the simulation. Consequently, the third column of each panel features black patches
representing these missing data regions\footnote{It is always possible to extract 
cuts at constant vertical height from the original snapshot that are not
affected by the missing regions, a direct consequence of the corrugated constant
optical depth surfaces.}. Interestingly, as long as at least one
sample from the posterior yields a valid value for a given pixel,
both the median and MAD can still be computed. In other words, the Bayesian
nature of the model allows us to extract meaningful information about the
physical parameters even for regions close to the limits of the line formation
region, so that there are fewer missing regions. This information is, however, highly uncertain, as indicated by the
increased MAD. Anyway, we checked that the estimated physical parameters are 
similar to those of the original snapshot at constant geometrical height.

Overall, the results show that the median model is highly similar to the depth stratification
of the physical quantities in the simulation. This is a remarkable finding,
demonstrating that the model can reliably learn the relationship between optical
depth and geometrical height by leveraging the correlations present in the
training data.

\subsection{Disambiguation and electric currents}
Since our model allows us to infer physical quantities in a geometrical scale as a natural consequence of taking 
advantage of the inherent statistical correlations present in the training data, it is almost straightforward to disambiguate
the transverse components of the magnetic field. Contrary to more standard methods, we do not need to rely on any field 
extrapolation method \citep[see, e.g.,][]{2006SoPh..237..267M} or extra assumptions \citep{2021A&A...647A.190B}. 
To this end, we use an in-house version of the 
minimum energy method developed by \cite{1994SoPh..155..235M} and later modified by \cite{2006SoPh..237..267M}
that is described in Appendix A. This makes it more similar to the method developed by \cite{2025ApJ...995..146Y}, which 
starts from an inverted stratified ambiguous field and is only based on the divergence-free nature of the magnetic field.

\begin{figure}
    \centering
    \includegraphics[width=\columnwidth]{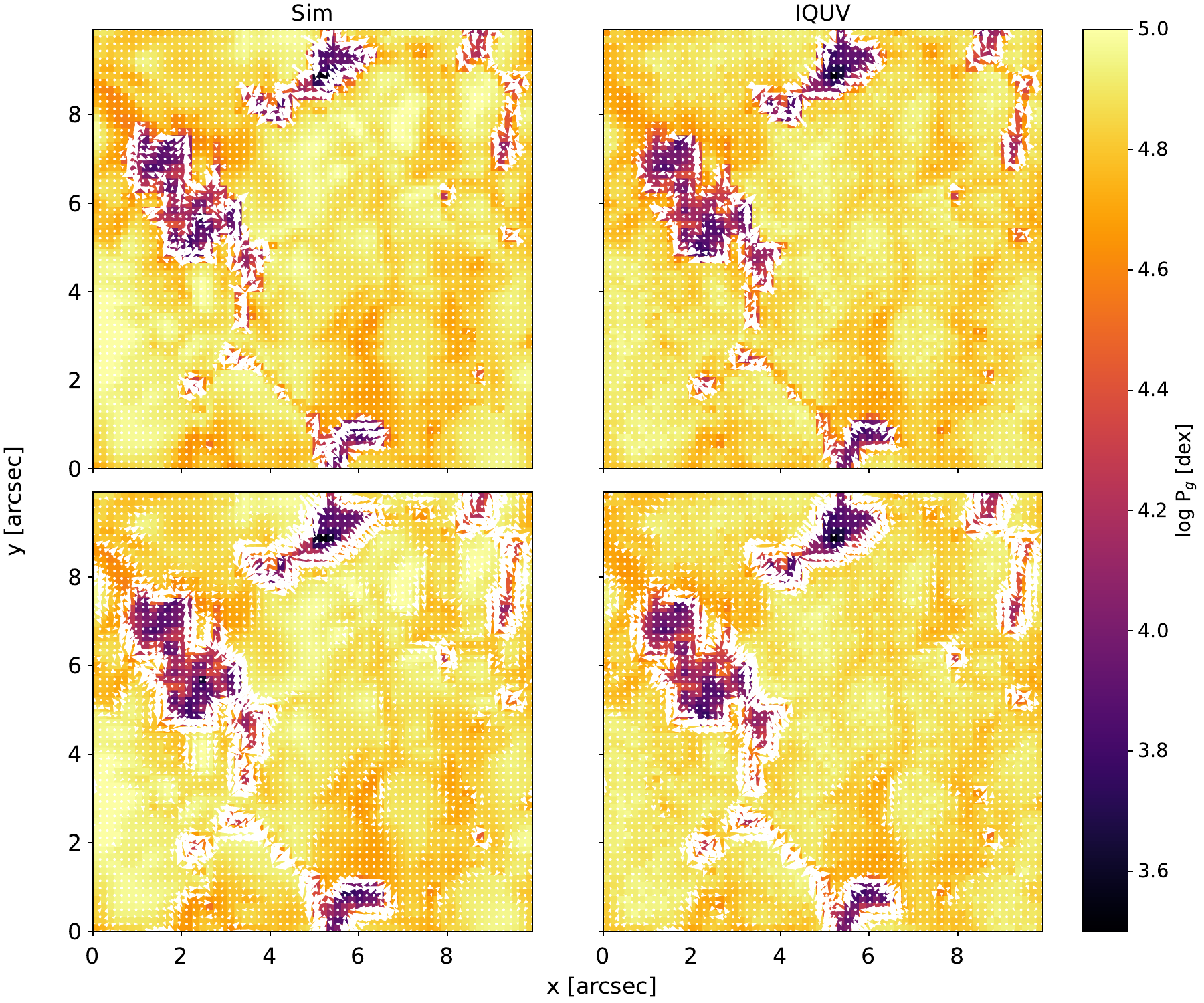}
    \caption{Logarithm of the gas pressure (background image), Lorentz force (arrows in the upper panels)
    and gas pressure gradient (arrows in the lower panels).\label{fig:validation_lorentz}}
\end{figure}

% During the optimization, the magnetic field divergence and the cartesian components of the 
% electric current density $\mathbf{J}$, given in A\,m$^{-2}$, are calculated using Ampere's law as:
% \begin{align}
% \nabla \cdot \mathbf{B} &= \frac{\partial B_x}{\partial x} + \frac{\partial B_y}{\partial y} + \frac{\partial B_z}{\partial z} \\
% J_x &= \mu_0^{-1} \left( \frac{\partial B_z}{\partial y} - \frac{\partial B_y}{\partial z} \right) \nonumber \\
% J_y &= \mu_0^{-1} \left( \frac{\partial B_x}{\partial z} - \frac{\partial B_z}{\partial x} \right) \nonumber \\
% J_z &= \mu_0^{-1} \left( \frac{\partial B_y}{\partial x} - \frac{\partial B_x}{\partial y} \right) \\
% \end{align}
% where $\mu_0=4\pi \times 10^{-7}$ kg\,m\,s$^{-2}$\,A$^{-2}$ is the magnetic permeability of free space
% and the magnetic fields are measured in T. The horizontal spatial derivatives are calculated using 
% finite differences. The vertical derivatives are calculated using the inferred atmospheric parameters at 
% different heights in the atmosphere. For this purpose, we fix 100 km as the target height for the disambiguation
% and then compute the vertical derivatives by interpolation at 90 and 100 km. 
Figure \ref{fig:validation_Bxyz} displays
the posterior median in the upper panel, the MAD in the middle panel, and the true 
values from the simulation in the lower panel for the ambiguous and disambiguated transverse components of the magnetic field. 
Although we can compute the criterion for the disambiguation at different heights, we choose to 
carry out the disambiguation at a fixed height of 100 km for visualization purposes.
The results demonstrate that computation of the vertical derivatives using the inferred height dependence is very
precise, producing disambiguated transverse components of the magnetic field that are strikingly similar to the true values from the simulation.
The posterior MAD for the more magnetized regions for $B_{p1}$ and $B_{p2}$ is smaller than in the rest of
the FoV, a consequence of the larger Stokes $Q$ and $U$ signals in those regions. The uncertainty in the
disambiguated transverse components of the magnetic field is, in general, larger than in the ambiguous components, 
which is specially relevant in the borders of the magnetized regions, where some patches show the wrong
polarity. However, all in all, the inferred model can be used to correctly disambiguate the more magnetized regions
of the quiet Sun, producing a solution with very low divergence and electric currents. 

\begin{figure}
    \centering
    \includegraphics[width=\columnwidth]{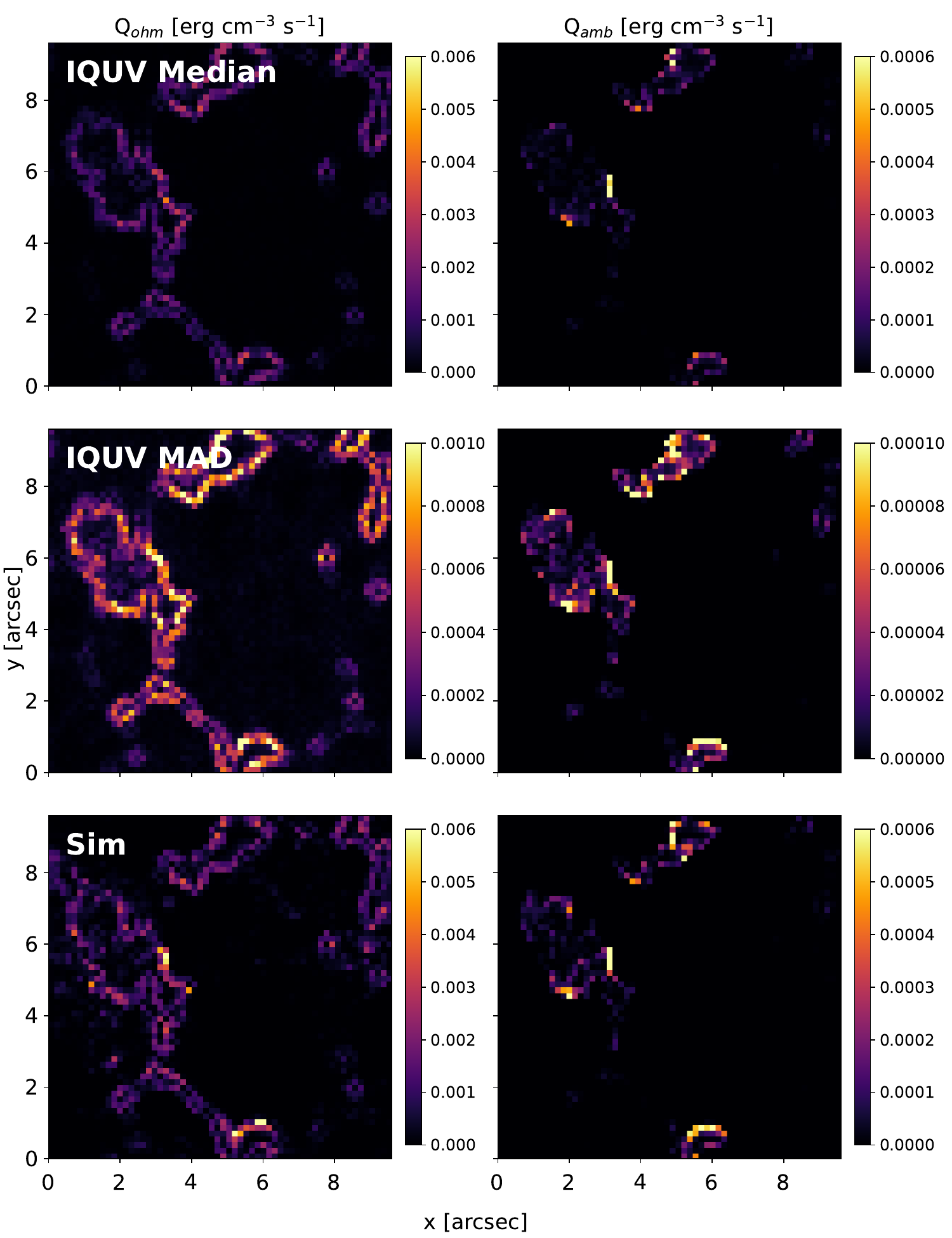}
    \caption{Ohmic (left) and ambipolar (right) dissipation maps for the validation set.\label{fig:validation_dissipation}}
\end{figure}

Having access to the disambiguated magnetic field allows us to calculate electric current density maps.
Electric currents are relevant for the energy balance of the solar atmosphere, since they can be dissipated 
by Ohmic and ambipolar diffusion \citep{1999Ap&SS.264...77P,2012ApJ...747...87K}. As such, they can 
be considered as proxies of highly non-potential magnetic fields, which can produce explosive
events in higher layers of the atmosphere. Since $\mathbf{J}$ depends on horizontal and vertical gradients
of the magnetic fields, they are often obtained from observations using magnetic field extrapolation methods 
or a-posteriori treatment of standard inversion methods \citep{2010ApJ...720.1417P,2010ApJ...721L..58P}.
However, some recent efforts \citep{2021A&A...656L..20P} have opened up the possibility of obtaining $\mathbf{J}$
from observations by resorting to the use of magnetohydrostatic equilibrium (MHS). These results
make use of a newly developed Stokes inversion code developed by \cite{2021A&A...647A.190B}, where
the MHS condition is imposed as a regularization during the inversion process. The method allows to
obtain the electric current density in the photosphere with sufficient precision, according to \cite{2023A&A...669A.122B}.
More recently, \cite{2025ApJ...995..146Y} demonstrated a new promising strategy that is based on 
using physics-informed neural networks to infer $\mathbf{J}$.

\begin{figure*}
    \centering
    \includegraphics[width=0.82\textwidth]{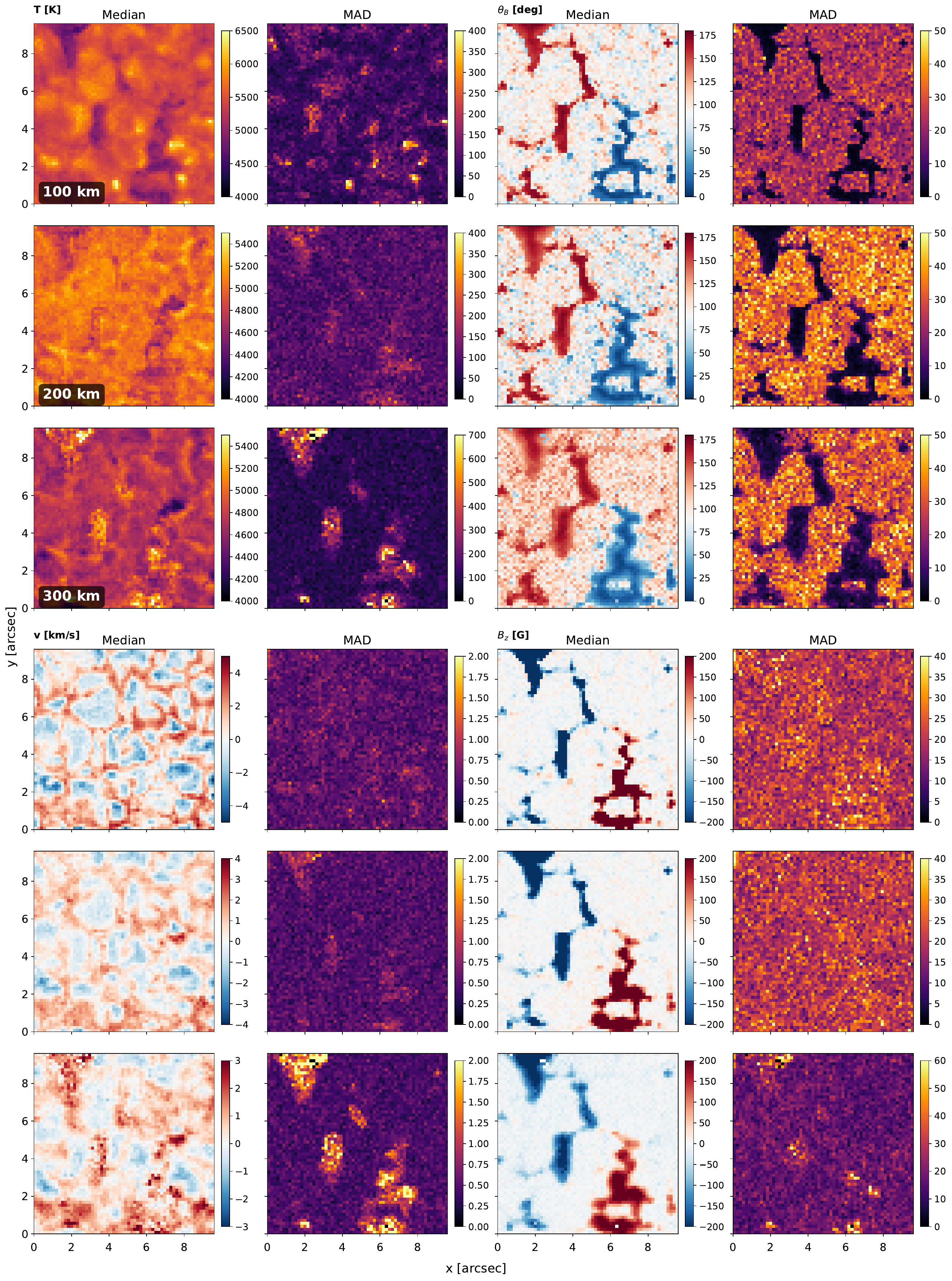}
    \caption{Inferred height-dependent temperature, line-of-sight velocity, inclination angle the magnetic
    field and line-of-sight component of the magnetic field for the quiet Sun Hinode/SP scan. Each variable 
    is shown at three different constant
    geometrical heights, with the two columns showing the median and the MAD of the distribution, respectively.\label{fig:hinode_maps}}
\end{figure*}

\begin{figure*}
    \centering
    \includegraphics[width=\textwidth]{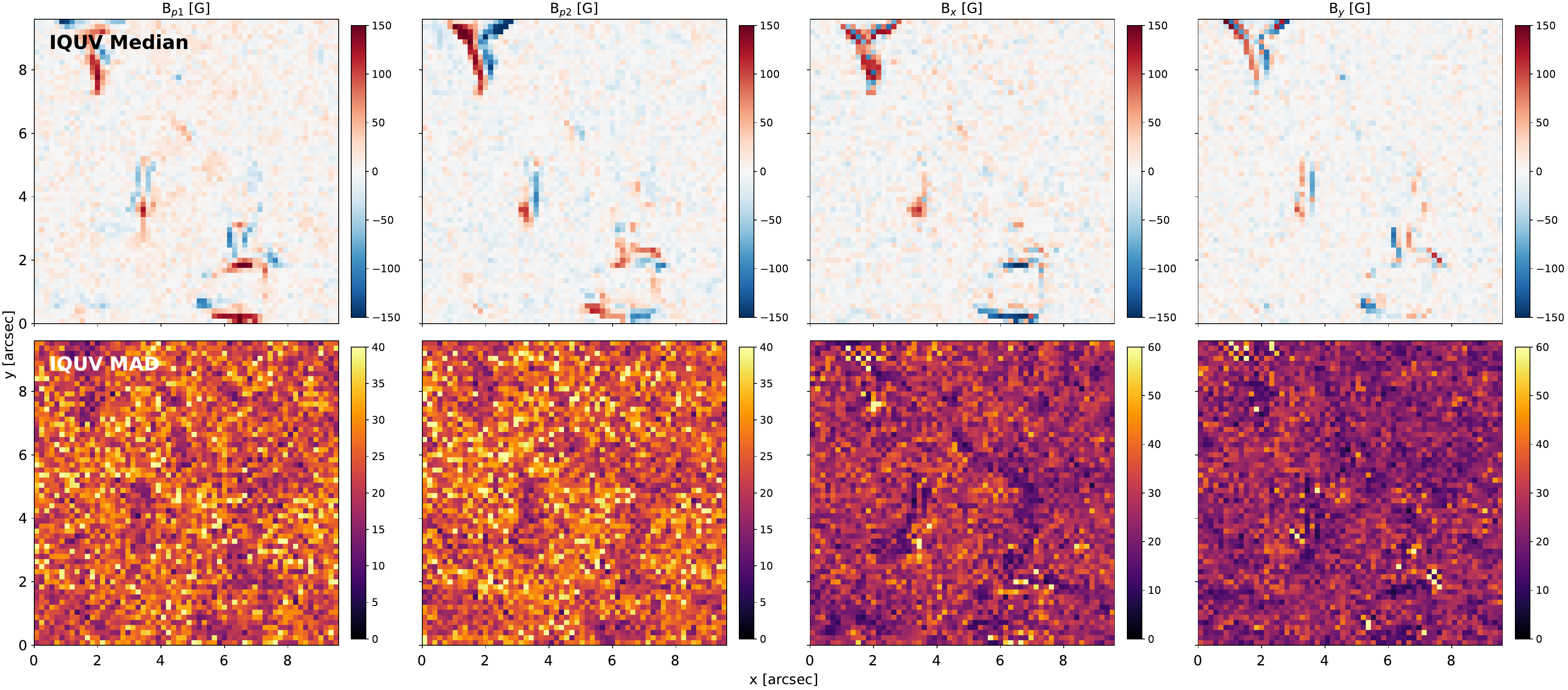}
    \caption{Inferred ambiguous (left two panels) and disambiguated (right two panels) transverse
    magnetic fields. The upper panels display the median of the posterior distribution, while the 
    lower panels show the MAD.\label{fig:hinode_Bxy}}
\end{figure*}

Contrary to previous approaches, in this work we can obtain $\mathbf{J}$ directly from 
the inferred magnetic field without any additional assumption,
apart from those encoded in the generative model. The resulting maps 
are displayed in Fig. \ref{fig:validation_currents}, with a direct comparison to the true values
obtained from the MHD snapshot. There is a remarkable similarity between the
median value and the value of $\mathbf{J}$ calculated from the true values of the simulation, despite
the increased uncertainty. It is worth noting that the value for $\mathbf{J}$ outside of clearly
magnetized regions is compatible with zero, with an uncertainty as low as 0.0025-0.0050 A\,m$^{-2}$.
In this sense, our result compares very favorable with those of \cite{2023A&A...669A.122B}, who
find a value of 0.05 A\,m$^{-2}$ for the regions where no polarimetric signal is detected.

\subsection{Lorentz force}
The Lorentz force $\mathbf{F}_L = \mathbf{J} \times \mathbf{B}$ can be computed
from the inferred electric current density and magnetic field. We can then compare its value 
with the inferred gas pressure gradient $\nabla P$ to check 
the consistency of the MHS equilibrium. Figure \ref{fig:validation_lorentz} shows maps of the gas pressure
at a geometrical height of 100 km as a background, both for the simulation (left panels) and the median value from the
posterior distribution (right panels).
The arrows in the upper panels show the horizontal components of the Lorentz force, computed from $\mathbf{J}$ 
and $\mathbf{B}$, while those in the lower panels show the horizontal components of the gas pressure gradient,
computed using finite differences. Their similarity is remarkable.

\subsection{Dissipation}
Finally, we can compute the Ohmic and ambipolar dissipation 
maps at any arbitrary height (note that the Hall dissipation is zero). 
They can be computed using \citep[see, e.g.,][]{2012ApJ...747...87K}:
\begin{equation}
Q_\mathrm{Ohm} = \eta J^2, \qquad Q_\mathrm{amb} = \eta_A \left| \mathbf{J} \times \mathbf{B} \right|^2,
\end{equation}
where
\begin{align}
\eta &= \frac{m_e \left( \nu_{ei} + \nu_{en} \right)}{e^2 n_e} \\
\eta_A &= \frac{\left( \rho_n / \rho \right)^2}{\rho_i \nu_{in} + \rho_e \nu_{en}}
\end{align}
are the Ohmic and ambipolar diffusivities, respectively. They depend on the collision frequencies between 
electrons, ions and neutrals, which are calculated using the formulas provided by \cite{2012ApJ...747...87K}:
\begin{align}
\nu_{in} &= n_n \sqrt{\frac{8 k_B T}{\pi m_{in}}} \sigma_{in} \nonumber \\
\nu_{en} &= n_n \sqrt{\frac{8 k_B T}{\pi m_{en}}} \sigma_{en} \nonumber \\
\nu_{ei} &= \frac{4}{3} \sqrt{\frac{2 \pi}{m_e}} \frac{n_e e^4 \Lambda}{(k_B T)^{3/2}}.
\end{align}
The cross sections take the values $\sigma_{in} = 5 \times 10^{-15}$ cm$^2$ and $\sigma_{en} = 10^{-15}$ cm$^2$, 
the Coulomb logarithm is calculated as $\Lambda = 23.4 - 1.15 \log_{10} n_e + 3.45 \log_{10} T$, where $n_e$ is 
measured in cm$^{-3}$ and $T$ in eV and $m_{in}$ and $m_{en}$ are the reduced masses of the ion-neutral and 
electron-neutral collisions, respectively. The density of neutrals and electrons is computed by
applying the LTE equation of state used in SIR to the inferred temperature and gas pressure, while that
of ions is obtained by applying the charge neutrality condition, so that 
$\rho_i = \rho_e$.

The resulting maps of $Q_\mathrm{Ohm}$ and $Q_\mathrm{amb}$ are shown in Fig. \ref{fig:validation_dissipation}, 
where we can see that the Ohmic dissipation is mostly localized in the borders of the magnetized 
regions, while the ambipolar dissipation is less extended and is only significant where there
is a measurable transverse component of the magnetic field. The inferred values of the dissipation are perfectly 
compatible with those obtained from the true values of the simulation. Uncertainties are larger in the
borders of the magnetized regions, but their amplitude are small, becoming only slightly larger than 10\%.

\section{Quiet Sun Hinode observations}
Having checked the performance of the model on synthetic data, we apply it to real observations. We use the 
Hinode/SP scan of a quiet Sun region at disk center taken on 2007-03-10 at 11:37 UT and analyzed in \cite{2008ApJ...672.1237L}.
The scan consists of a 1024$\times$2048 pixels with a spatial sampling of 0.16" per pixel, of which we extract
a tiny subfield of 64$\times$64 pixels that contains magnetic elements of the two polarities.
The Stokes profiles are normalized in the same way as the training data, and then 
fed into the conditioning U-Net to extract the latent representation. The flow matching model then produces 
25 samples of the posterior distribution.

\begin{figure*}
    \centering
    \includegraphics[width=\textwidth]{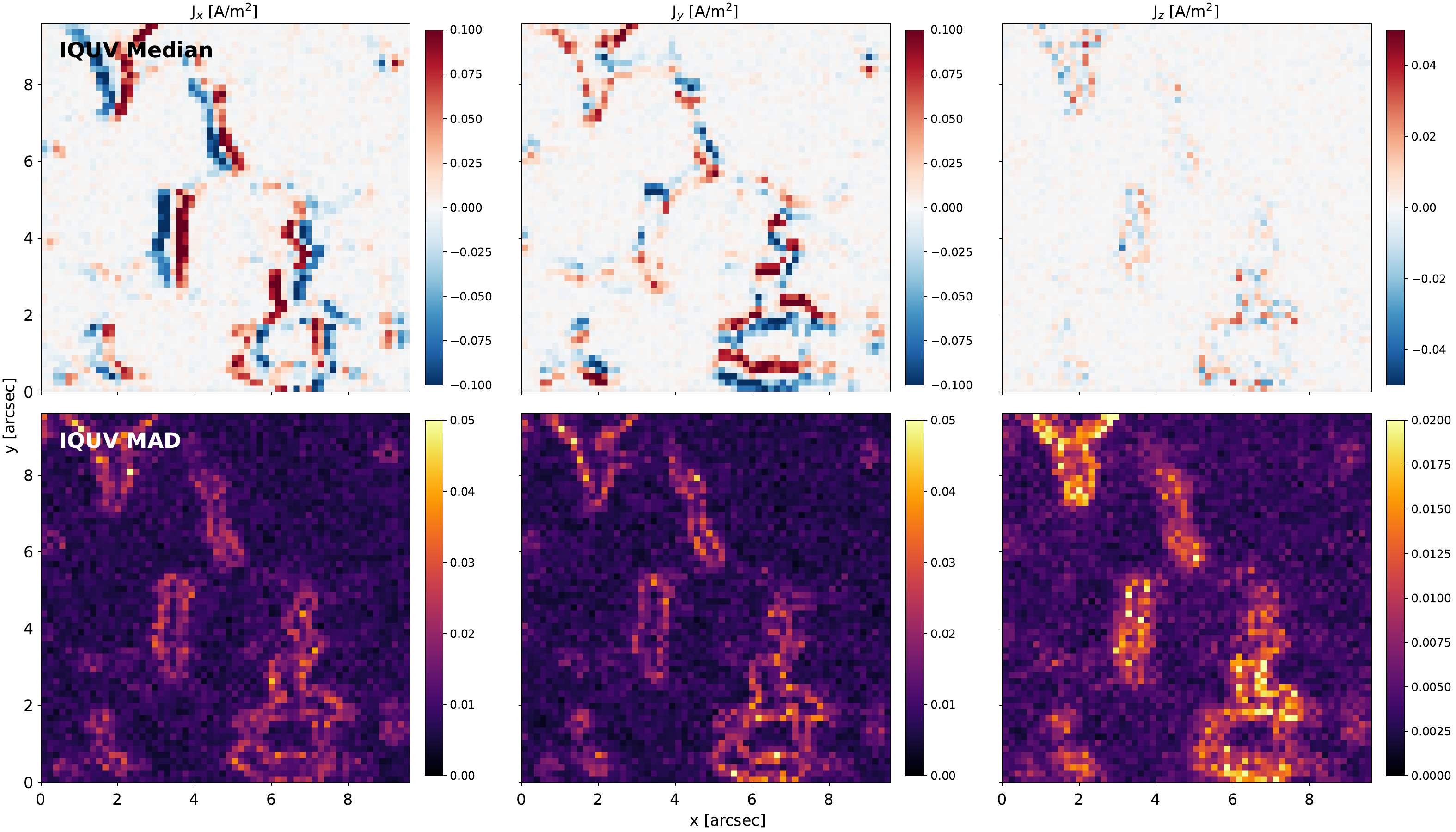}
    \caption{Inferred electric current density maps for the Hinode/SP quiet Sun observation of Fig. \ref{fig:hinode_maps}.\label{fig:hinode_currents}}
\end{figure*}
\begin{figure}
    \centering
    \includegraphics[width=\columnwidth]{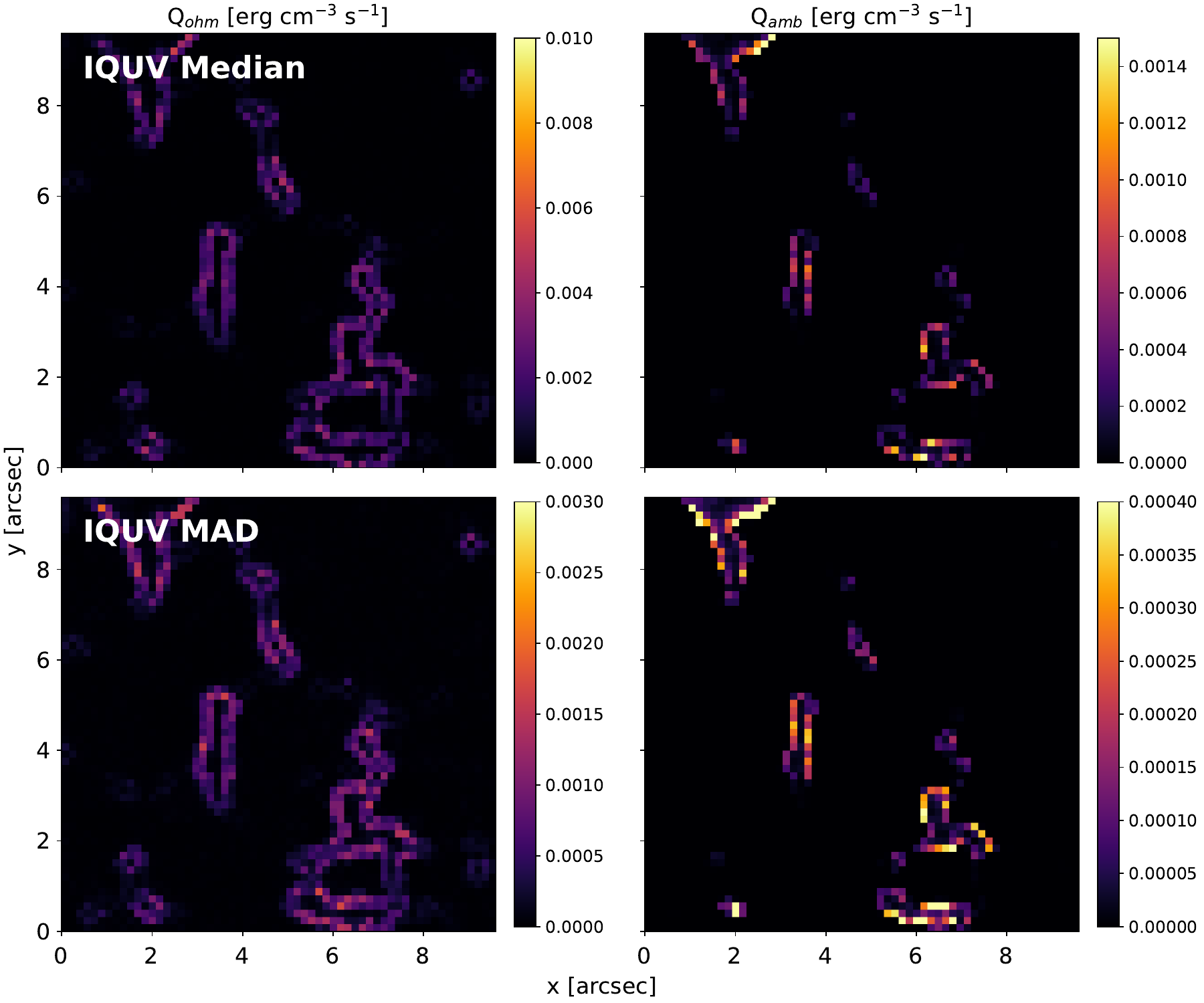}
    \caption{Ohmic (left) and ambipolar (right) dissipation maps for the Hinode/SP quiet Sun 
    observation of Fig. \ref{fig:hinode_maps}.\label{fig:hinode_dissipation}}
\end{figure}

The resulting maps at constant geometrical height for $T$, $v$, $B_z$, and $\theta_B$ are shown in 
Fig. \ref{fig:hinode_maps}, where we show the median and the MAD of the posterior distribution. Given that
the flow matching model produces samples from the posterior, derived quantities like $\theta_B$ are trivially obtained
by calculating the derived quantity for each sample and then calculating the statistics of the resulting distribution.
The results show that the magnetized regions contain vertical magnetic fields of both polarities at the
core of the magnetic elements, with a strong horizontal gradient towards essentially undetectable magnetic
fields in the granular regions. The uncertainty of the model on the inclination of the field in the 
magnetized regions is almost negligible. We also witness the canopy structure of the magnetic
field, with broader structures at 300 km than at 100 km. 

Motivated by the encouraging results obtained in the validation set concerning the estimation of
$\mathbf{J}$, we apply the minimum energy method to disambiguate the transverse
components of the magnetic field to compute such current densities. The disambiguated transverse
components are shown in Fig. \ref{fig:hinode_Bxy}, where we show the ambiguous and disambiguated transverse 
components of the magnetic field, together with their MAD. 
Except for some very localized patches in which the MAD is sizable, the results seem to indicate 
a very clear disambiguated solution. The majority of the FoV is compatible with 
zero transverse magnetic field, which is expected in the quiet Sun, given that the Stokes $Q$ and $U$ signals 
are very weak. Despite that the only patches with significant transverse magnetic field are those 
corresponding to the strongest magnetic elements displayed in Fig. \ref{fig:hinode_maps}, the 
inferred $\mathbf{J}$, shown in Fig. \ref{fig:hinode_currents}, is very well localized. Strong currents appear,
with both signs, in the borders of the magnetized regions. The posterior uncertainty is sufficiently
low to ensure that the measured currents are real and not a consequence of the noise in the data.

Equipped with the inferred $\mathbf{B}$ and $\mathbf{J}$, we can calculate the Ohmic and ambipolar dissipation 
maps, which are shown in Fig. \ref{fig:hinode_dissipation}. The results show that the both types of 
dissipation are compatible with zero in the regions where we do not detect any magnetic field. Again, 
Ohmic dissipation is mostly localized in the borders of the magnetized regions, while the ambipolar
dissipation is more important in regions with a measurable transverse component of the magnetic field, and
much less spatially extended. Ambipolar dissipation is also smaller in absolute value than Ohmic dissipation. This
is expected in the photosphere, where the plasma is mostly neutral and the magnetic field is not strong enough to 
strongly decouple neutrals and ions.

\begin{figure*}
    \centering
    \includegraphics[width=0.73\textwidth]{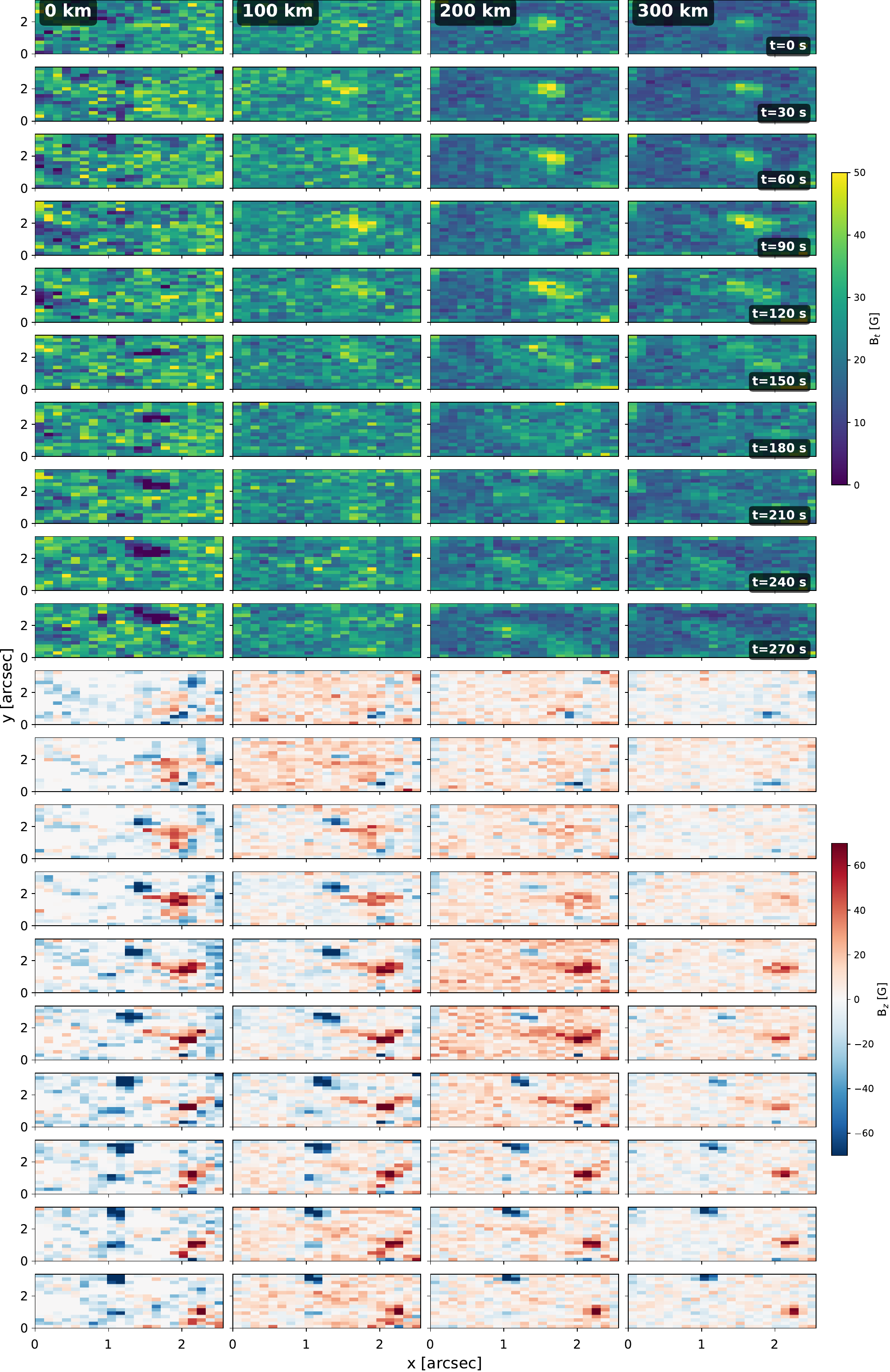}
    \caption{Transverse (upper panels) and longitudinal (lower panels) components of the magnetic field for the 
    emerging small-scale loop.\label{fig:hinode_loop}}
\end{figure*}

\section{Small-scale loop emergence}
As a final application, we show a preliminary analysis of the Hinode/SP scans of one
of the small-scale loop emergence events analyzed by \cite{2009ApJ...700.1391M}. A more detailed analysis will be presented in a future study. 

Small-scale magnetic loops were identified as two opposite-polarity circular polarization 
signals flanking a linear polarization signal \citep{MartinezGonzalez2007, Centeno2007}. Their 
temporal coherence strongly suggested that they corresponded to organized, connected magnetic 
field structures in the form of $\Omega$- or U-loops. Their three-dimensional geometry was 
previously inferred by \cite{Ishikawa2010} and \cite{MartinezGonzalez2010} through pixel-by-pixel 
inversions in the optical depth scale and in a non-consistent geometrical height, respectively. 
Here, we reconstruct the three-dimensional geometry in a consistent geometrical height scale and analyze 
the temporal evolution of the transverse and longitudinal components of the magnetic field as a 
function of geometrical height.

In Figure \ref{fig:hinode_loop}, a three-dimensional magnetic structure in the form of a loop 
ascending through the photosphere can be clearly identified. During the first two snapshots, only the 
transverse magnetic field is visible, corresponding to the apex of the loop. This predominantly 
horizontal structure exhibits an approximate vertical extent of 200 km, with the transverse field 
reaching its maximum intensity around that height. 

At t=60 s, opposite-polarity regions appear at both ends of the transverse field, extending
from 0 to 100 km. These regions correspond to the loop footpoints connecting to the apex. At a height 
of 200 km, the same polarity pattern and a slight inclination of the footpoints, reflected in a 
weaker longitudinal field component, are observed, supporting the interpretation of a single 
connected magnetic structure. At t=90 s, the footpoints become visible up to 300 km, while the 
linear polarization signal is strongest at the highest layers. This behavior indicates that the 
loop continues to rise through the photosphere.

At t=120 s, the transverse field weakens and the longitudinal field at the footpoints becomes 
stronger at all heights. This supports the hypothesis of an ascending loop, since the upper part 
of the loop is already leaving the line formation region and the more vertical part of the loop 
is penetrating this region. At this stage, the loop reaches a height of approximately 300 km and 
a horizontal extent of about 700 km, resulting in a very flattened loop. This is consistent with 
the results of \cite{MartinezGonzalez2010}, who, for t=120 s, reported a loop height of 300 km and 
a horizontal extent of about 750 km (see their Figure 1). Finally, the apex of the loop leaves the 
formation region at t=150 s, and only the footpoints remain visible as they continue to separate with time.

%As a final application of the newly developed code, we show a preliminary analysis of the Hinode/SP scans of one
%of the small-scale loop emergence events analyzed by \cite{2009ApJ...700.1391M}. The full 
%analysis will be presented in a more detailed analysis in a future work. Given the ability of the
%code to infer the geometrical height dependence of the atmospheric parameters, we can analyze the 
%temporal evolution of the transverse and longitudinal components of the magnetic field. Figure \ref{fig:hinode_loop}
%shows the time evolution of the inferred magnetic field during 4.5 min when one of the small-scale
%loops is emerging. The results clearly demonstrate that the top of the loop appears earlier in the upper
%layers in $B_t$, showing signatures at heights of 200-300 km at $t=30$ s. The feet of the loop (producing
%a signature in $B_z$) are only visible at very deep layers initially, and only appear at 200 km after 90 s 
%and at 300 km after 120-150 s. This is a clear demonstration of the loop-like structure emerging from the
%deeper layers.

\section{Conclusions and future work}
We have presented \code, a new code for the inversion of Stokes profiles based on conditional 
flow matching. The code has been trained with the quiet Sun simulations of the SPIN4D project, for
the pair of lines at 630 nm and for disk center observations. It has been validated with a 
snapshot with mixed polarities that was not used during training. The results show that the 
model is able to capture the true values of the atmospheric parameters, providing samples from the posterior
distribution, which allows us to characterize the uncertainty and correlation of the inferred parameters.
The model is able to return the physical parameters as a function of geometrical depth, which
allows us to disambiguate the transverse components of the magnetic field without any 
field extrapolation method. This opens up the possibility of calculating electric current 
density maps and dissipation with high precision. 

The application to Hinode/SP quiet Sun observations shows that the model can be safely applied to real data.
We have produced, for the first time, reliable (with uncertainties associated) maps of the electric current density
and Ohmic and ambipolar dissipation in the quiet Sun. Additionally, we find unequivocal 
evidence of the loop-like structure in geometrical height of the magnetic field.

We are perfectly aware of the limitations of this code and we have long-term ideas to address them. First, 
the results are dependent on the availability and realism of MHD simulations. For this reason, only quiet Sun 
simulations in the photosphere have been used for training, where the state-of-the-art MHD simulations are 
considered to be sufficiently realistic. For this reason, the model is expected to perform well 
only in quiet Sun regions in the photosphere. More active regions and higher layers of the atmosphere are not included in 
the training data because of the lack of consensus on how realistic these MHD simulations are. The model can be trivially 
extended to those regions once such simulations are available. 

Second, the model has been trained with disk center observations at the pair of Fe \textsc{i} lines
at 630 nm. Extending it to other lines of sight is straightforward, although we would need to add 
a conditional variable to the model to encode the angle of observation. Additionally, alternative 
encoders of the Stokes profiles (e.g., based on the transformer architecture) could be used to make 
the model more flexible and able to deal with different spectral lines.

Third, we have limited the training to the relatively reduced spatial resolution provided by the Hinode/SP instrument.
On one hand, this allows the model to be applied to the enormous amount of Hinode/SP data available in the archives. On 
the other hand, it limits the performance of the model for more modern telescopes, such as 
the Daniel K. Inouye Solar Telescope \citep[DKIST;][]{2020SoPh..295..172R} and the future European Solar 
Telescope \citep[EST;][]{2022A&A...666A..21Q}. The extension to higher spatial resolution is feasible but not
straightforward. From a computational point of view, the transfer and storage of MHD simulations at a resolution of 28 km (the DKIST 
diffraction limit at 630 nm) for training is difficult. On the other hand, training generative models directly
in pixel space at such high resolution is probably not feasible. For this reason, we speculate that the extension to 
higher spatial resolution will require the use of a generative flow matching model in a 
latent representation. Following \cite{rombach2022high}, this latent representation can be obtained
by pre-training a variational autoencoder (VAE) on the MHD simulations, and then 
training the flow matching model in the latent space of the VAE.

Additionally, in the present implementation, the azimuthal disambiguation is performed as a separate 
post-processing step. Since the adopted algorithm makes use of information from neighbouring pixels, it 
is possible to incorporate the disambiguation directly into the Stokes inversion through a spatially 
coupled convolutional neural network-based framework. This extension is currently in preparation \citep{Nick2026}.

Finally, the current version of the code is not providing any information about the transverse velocity components, 
which are relevant for the energy balance of the solar atmosphere. Although they can be easily estimated from 
time series of observations \citep{2017A&A...604A..11A}, we want to explore whether it is possible to
infer them by exploiting the correlation between the velocity and magnetic field in the training data.

\begin{acknowledgements}
We acknowledge the critical read of an early draft provided by J. de la Cruz Rodr\'{\i}guez.
A.A.R. acknowledges funding from the Agencia Estatal de Investigación del Ministerio de Ciencia, Innovación y Universidades (MCIU/AEI) under grant 
``Polarimetric Inference of Magnetic Fields'' and the European Regional Development Fund (ERDF) with reference PID2022-136563NB-I00/10.13039/501100011033.
K.Y., S.C.D, and X.S. are supported by award 2008334 from the US National Science Foundation.
This research has made use of NASA's Astrophysics Data System Bibliographic Services.
The code is publicly available at \urlcode.
We acknowledge the community effort devoted to the development of the following open-source packages that were
used in this work: \texttt{numpy} \citep[\texttt{numpy.org},][]{numpy20}, 
\texttt{matplotlib} \citep[\texttt{matplotlib.org},][]{matplotlib}, \texttt{PyTorch} 
\citep[\texttt{pytorch.org},][]{pytorch19}, \texttt{scipy} \citep[\texttt{scipy.org},][]{2020SciPy-NMeth} and \texttt{scikit-learn} \citep[\texttt{scikit-learn.org},][]{scikit-learn}.
\end{acknowledgements}

\bibliographystyle{aa} % style aa.bst
\bibliography{biblio} % your references Yourfile.bib

@PROCEEDINGS{2004ASSL..307.....L,
    title = "{Polarization in Spectral Lines}",
booktitle = {Astrophysics and Space Science Library},
     year = 2004,
   series = {Astrophysics and Space Science Library},
   volume = 307,
   editor = {{Landi Degl'Innocenti}, E. and {Landolfi}, M.},
    month = aug,
      doi = {10.1007/978-1-4020-2415-3},
   adsurl = {http://adsabs.harvard.edu/abs/2004ASSL..307.....L}
}

@ARTICLE{2015A&A...577A...7S,
   author = {{Socas-Navarro}, H. and {de la Cruz Rodr{\'{\i}}guez}, J. and 
	{Asensio Ramos}, A. and {Trujillo Bueno}, J. and {Ruiz Cobo}, B.
	},
    title = "{An open-source, massively parallel code for non-LTE synthesis and inversion of spectral lines and Zeeman-induced Stokes profiles}",
  journal = {\aap},
archivePrefix = "arXiv",
   eprint = {1408.6101},
 primaryClass = "astro-ph.SR",
     year = 2015,
    month = may,
   volume = 577,
      eid = {A7},
    pages = {A7},
      doi = {10.1051/0004-6361/201424860},
   adsurl = {http://adsabs.harvard.edu/abs/2015A%26A...577A...7S}
}

@ARTICLE{2008SoPh..249..167T,
   author = {{Tsuneta}, S. and {Ichimoto}, K. and {Katsukawa}, Y. and {Nagata}, S. and 
	{Otsubo}, M. and {Shimizu}, T. and {Suematsu}, Y. and {Nakagiri}, M. and 
	{Noguchi}, M. and {Tarbell}, T. and {Title}, A. and {Shine}, R. and 
	{Rosenberg}, W. and {Hoffmann}, C. and {Jurcevich}, B. and {Kushner}, G. and 
	{Levay}, M. and {Lites}, B. and {Elmore}, D. and {Matsushita}, T. and 
	{Kawaguchi}, N. and {Saito}, H. and {Mikami}, I. and {Hill}, L.~D. and 
	{Owens}, J.~K.},
    title = "{The Solar Optical Telescope for the Hinode Mission: An Overview}",
  journal = {\solphys},
archivePrefix = "arXiv",
   eprint = {0711.1715},
     year = 2008,
    month = jun,
   volume = 249,
    pages = {167-196},
      doi = {10.1007/s11207-008-9174-z},
   adsurl = {http://adsabs.harvard.edu/abs/2008SoPh..249..167T}
}

@ARTICLE{2009ApJ...691..640R,
   author = {{Rempel}, M. and {Sch{\"u}ssler}, M. and {Kn{\"o}lker}, M.},
    title = "{Radiative Magnetohydrodynamic Simulation of Sunspot Structure}",
  journal = {\apj},
archivePrefix = "arXiv",
   eprint = {0808.3294},
     year = 2009,
    month = jan,
   volume = 691,
    pages = {640-649},
      doi = {10.1088/0004-637X/691/1/640},
   adsurl = {http://adsabs.harvard.edu/abs/2009ApJ...691..640R}
}

@ARTICLE{2014ApJ...789..132R,
   author = {{Rempel}, M.},
    title = "{Numerical Simulations of Quiet Sun Magnetism: On the Contribution from a Small-scale Dynamo}",
  journal = {\apj},
archivePrefix = "arXiv",
   eprint = {1405.6814},
 primaryClass = "astro-ph.SR",
     year = 2014,
    month = jul,
   volume = 789,
      eid = {132},
    pages = {132},
      doi = {10.1088/0004-637X/789/2/132},
   adsurl = {http://adsabs.harvard.edu/abs/2014ApJ...789..132R}
}

@ARTICLE{2005A&A...429..335V,
   author = {{V{\"o}gler}, A. and {Shelyag}, S. and {Sch{\"u}ssler}, M. and 
	{Cattaneo}, F. and {Emonet}, T. and {Linde}, T.},
    title = "{Simulations of magneto-convection in the solar photosphere.  Equations, methods, and results of the MURaM code}",
  journal = {\aap},
     year = 2005,
    month = jan,
   volume = 429,
    pages = {335-351},
      doi = {10.1051/0004-6361:20041507},
   adsurl = {http://adsabs.harvard.edu/abs/2005A%26A...429..335V}
}

@ARTICLE{1992ApJ...398..375R,
   author = {{Ruiz Cobo}, B. and {del Toro Iniesta}, J.~C.},
    title = "{Inversion of Stokes profiles}",
  journal = {\apj},
     year = 1992,
    month = oct,
   volume = 398,
    pages = {375-385},
      doi = {10.1086/171862},
   adsurl = {http://adsabs.harvard.edu/abs/1992ApJ...398..375R}
}

@MANUAL{NICOLE,
  title = {NICOLE Non-LTE Inversion COde using the Lorien Engine},
  author = {{Socas-Navarro}, H., {de la Cruz}, J. and {Asensio-Ramos}, A.},
  year = "version 14.07"
   }

@ARTICLE{2008A&A...484L..17D,
   author = {{Danilovic}, S. and {Gandorfer}, A. and {Lagg}, A. and {Sch{\"u}ssler}, M. and 
	{Solanki}, S.~K. and {V{\"o}gler}, A. and {Katsukawa}, Y. and 
	{Tsuneta}, S.},
    title = "{The intensity contrast of solar granulation: comparing Hinode SP results with MHD simulations}",
  journal = {\aap},
archivePrefix = "arXiv",
   eprint = {0804.4230},
     year = 2008,
    month = jun,
   volume = 484,
    pages = {L17-L20},
      doi = {10.1051/0004-6361:200809857},
   adsurl = {http://adsabs.harvard.edu/abs/2008A%26A...484L..17D}
}

@incollection{pytorch19,
title = {PyTorch: An Imperative Style, High-Performance Deep Learning Library},
author = {Paszke, Adam and Gross, Sam and Massa, Francisco and Lerer, Adam and Bradbury, James and Chanan, Gregory and Killeen, Trevor and Lin, Zeming and Gimelshein, Natalia and Antiga, Luca and Desmaison, Alban and Kopf, Andreas and Yang, Edward and DeVito, Zachary and Raison, Martin and Tejani, Alykhan and Chilamkurthy, Sasank and Steiner, Benoit and Fang, Lu and Bai, Junjie and Chintala, Soumith},
booktitle = {Advances in Neural Information Processing Systems 32},
editor = {H. Wallach and H. Larochelle and A. Beygelzimer and F. d'\'{e}-Buc and E. Fox and R. Garnett},
pages = {8024--8035},
year = {2019},
publisher = {Curran Associates, Inc.},
url = {http://papers.neurips.cc/paper/9015-pytorch-an-imperative-style-high-performance-deep-learning-library.pdf}
}

@inproceedings{adam14,
  author = {Kingma, Diederik P. and Ba, Jimmy},
  booktitle = {ICLR (Poster)},  
  editor = {Bengio, Yoshua and LeCun, Yann},
  ee = {http://arxiv.org/abs/1412.6980},
  title = {Adam: A Method for Stochastic Optimization.},
  url = {http://dblp.uni-trier.de/db/conf/iclr/iclr2015.html#KingmaB14},
  year = 2015
}

@ARTICLE{numpy20,
  author  = {Harris, Charles R. and Millman, K. Jarrod and
            van der Walt, Stéfan J and Gommers, Ralf and
            Virtanen, Pauli and Cournapeau, David and
            Wieser, Eric and Taylor, Julian and Berg, Sebastian and
            Smith, Nathaniel J. and Kern, Robert and Picus, Matti and
            Hoyer, Stephan and van Kerkwijk, Marten H. and
            Brett, Matthew and Haldane, Allan and
            Fernández del Río, Jaime and Wiebe, Mark and
            Peterson, Pearu and Gérard-Marchant, Pierre and
            Sheppard, Kevin and Reddy, Tyler and Weckesser, Warren and
            Abbasi, Hameer and Gohlke, Christoph and
            Oliphant, Travis E.},
  title   = {Array programming with {NumPy}},
  journal = {Nature},
  year    = {2020},
  volume  = {585},
  pages   = {357–362},
  doi     = {10.1038/s41586-020-2649-2}
}

@ARTICLE{matplotlib,
  author={J. D. {Hunter}},
  journal={Computing in Science   Engineering}, 
  title={Matplotlib: A 2D Graphics Environment}, 
  year={2007},
  volume={9},
  number={3},
  pages={90-95},}

@ARTICLE{Ishikawa2010,
       author = {{Ishikawa}, Ryohko and {Tsuneta}, Saku and {Jur{\v{c}}{\'a}k}, Jan},
        title = "{Three-Dimensional View of Transient Horizontal Magnetic Fields in the Photosphere}",
      journal = {\apj},
         year = 2010,
        month = apr,
       volume = {713},
       number = {2},
        pages = {1310-1321},
          doi = {10.1088/0004-637X/713/2/1310},
archivePrefix = {arXiv},
       eprint = {1003.1376},
 primaryClass = {astro-ph.SR},
       adsurl = {https://ui.adsabs.harvard.edu/abs/2010ApJ...713.1310I}
}

@ARTICLE{Centeno2007,
       author = {{Centeno}, R. and {Socas-Navarro}, H. and {Lites}, B. and {Kubo}, M. and {Frank}, Z. and {Shine}, R. and {Tarbell}, T. and {Title}, A. and {Ichimoto}, K. and {Tsuneta}, S. and {Katsukawa}, Y. and {Suematsu}, Y. and {Shimizu}, T. and {Nagata}, S.},
        title = "{Emergence of Small-Scale Magnetic Loops in the Quiet-Sun Internetwork}",
      journal = {\apjl},
         year = 2007,
        month = sep,
       volume = {666},
       number = {2},
        pages = {L137-L140},
          doi = {10.1086/521726},
archivePrefix = {arXiv},
       eprint = {0708.0844},
 primaryClass = {astro-ph},
       adsurl = {https://ui.adsabs.harvard.edu/abs/2007ApJ...666L.137C}
}

@ARTICLE{MartinezGonzalez2007,
       author = {{Mart{\'\i}nez Gonz{\'a}lez}, M.~J. and {Collados}, M. and {Ruiz Cobo}, B. and {Solanki}, S.~K.},
        title = "{Low-lying magnetic loops in the solar internetwork}",
      journal = {\aap},
         year = 2007,
        month = jul,
       volume = {469},
       number = {3},
        pages = {L39-L42},
          doi = {10.1051/0004-6361:20077505},
archivePrefix = {arXiv},
       eprint = {0705.1319},
 primaryClass = {astro-ph},
       adsurl = {https://ui.adsabs.harvard.edu/abs/2007A&A...469L..39M}
}

@ARTICLE{MartinezGonzalez2010,
       author = {{Mart{\'\i}nez Gonz{\'a}lez}, M.~J. and {Manso Sainz}, R. and {Asensio Ramos}, A. and {Bellot Rubio}, L.~R.},
        title = "{Small Magnetic Loops Connecting the Quiet Surface and the Hot Outer Atmosphere of the Sun}",
      journal = {\apjl},
         year = 2010,
        month = may,
       volume = {714},
       number = {1},
        pages = {L94-L97},
          doi = {10.1088/2041-8205/714/1/L94},
archivePrefix = {arXiv},
       eprint = {1003.1255},
 primaryClass = {astro-ph.SR},
       adsurl = {https://ui.adsabs.harvard.edu/abs/2010ApJ...714L..94M}
}

@article{scikit-learn,
  title={Scikit-learn: Machine Learning in {P}ython},
  author={Pedregosa, F. and Varoquaux, G. and Gramfort, A. and Michel, V.
          and Thirion, B. and Grisel, O. and Blondel, M. and Prettenhofer, P.
          and Weiss, R. and Dubourg, V. and Vanderplas, J. and Passos, A. and
          Cournapeau, D. and Brucher, M. and Perrot, M. and Duchesnay, E.},
  journal={Journal of Machine Learning Research},
  volume={12},
  pages={2825--2830},
  year={2011}
}

@ARTICLE{2020SciPy-NMeth,
  author  = {Virtanen, Pauli and Gommers, Ralf and Oliphant, Travis E. and
            Haberland, Matt and Reddy, Tyler and Cournapeau, David and
            Burovski, Evgeni and Peterson, Pearu and Weckesser, Warren and
            Bright, Jonathan and {van der Walt}, St{\'e}fan J. and
            Brett, Matthew and Wilson, Joshua and Millman, K. Jarrod and
            Mayorov, Nikolay and Nelson, Andrew R. J. and Jones, Eric and
            Kern, Robert and Larson, Eric and Carey, C J and
            Polat, {\.I}lhan and Feng, Yu and Moore, Eric W. and
            {VanderPlas}, Jake and Laxalde, Denis and Perktold, Josef and
            Cimrman, Robert and Henriksen, Ian and Quintero, E. A. and
            Harris, Charles R. and Archibald, Anne M. and
            Ribeiro, Ant{\^o}nio H. and Pedregosa, Fabian and
            {van Mulbregt}, Paul and {SciPy 1.0 Contributors}},
  title   = {{{SciPy} 1.0: Fundamental Algorithms for Scientific
            Computing in Python}},
  journal = {Nature Methods},
  year    = {2020},
  volume  = {17},
  pages   = {261--272},
  adsurl  = {https://rdcu.be/b08Wh},
  doi     = {10.1038/s41592-019-0686-2},
}

@ARTICLE{2020SoPh..295..172R,
       author = {{Rimmele}, Thomas R. and {Warner}, Mark and {Keil}, Stephen L. and {Goode}, Philip R. and {Kn{\"o}lker}, Michael and {Kuhn}, Jeffrey R. and {Rosner}, Robert R. and {McMullin}, Joseph P. and {Casini}, Roberto and {Lin}, Haosheng and {W{\"o}ger}, Friedrich and {von der L{\"u}he}, Oskar and {Tritschler}, Alexandra and {Davey}, Alisdair and {de Wijn}, Alfred and {Elmore}, David F. and {Fehlmann}, Andr{\'e} and {Harrington}, David M. and {Jaeggli}, Sarah A. and {Rast}, Mark P. and {Schad}, Thomas A. and {Schmidt}, Wolfgang and {Mathioudakis}, Mihalis and {Mickey}, Donald L. and {Anan}, Tetsu and {Beck}, Christian and {Marshall}, Heather K. and {Jeffers}, Paul F. and {Oschmann}, Jacobus M. and {Beard}, Andrew and {Berst}, David C. and {Cowan}, Bruce A. and {Craig}, Simon C. and {Cross}, Eric and {Cummings}, Bryan K. and {Donnelly}, Colleen and {de Vanssay}, Jean-Benoit and {Eigenbrot}, Arthur D. and {Ferayorni}, Andrew and {Foster}, Christopher and {Galapon}, Chriselle Ann and {Gedrites}, Christopher and {Gonzales}, Kerry and {Goodrich}, Bret D. and {Gregory}, Brian S. and {Guzman}, Stephanie S. and {Guzzo}, Stephen and {Hegwer}, Steve and {Hubbard}, Robert P. and {Hubbard}, John R. and {Johansson}, Erik M. and {Johnson}, Luke C. and {Liang}, Chen and {Liang}, Mary and {McQuillen}, Isaac and {Mayer}, Christopher and {Newman}, Karl and {Onodera}, Brialyn and {Phelps}, LeEllen and {Puentes}, Myles M. and {Richards}, Christopher and {Rimmele}, Lukas M. and {Sekulic}, Predrag and {Shimko}, Stephan R. and {Simison}, Brett E. and {Smith}, Brett and {Starman}, Erik and {Sueoka}, Stacey R. and {Summers}, Richard T. and {Szabo}, Aimee and {Szabo}, Louis and {Wampler}, Stephen B. and {Williams}, Timothy R. and {White}, Charles},
        title = "{The Daniel K. Inouye Solar Telescope - Observatory Overview}",
      journal = {\solphys},
         year = 2020,
        month = dec,
       volume = {295},
       number = {12},
          eid = {172},
        pages = {172},
          doi = {10.1007/s11207-020-01736-7},
       adsurl = {https://ui.adsabs.harvard.edu/abs/2020SoPh..295..172R}
}

@ARTICLE{1977SoPh...55...47A,
       author = {{Auer}, L.~H. and {Heasley}, J.~N. and {House}, L.~L.},
        title = "{The determination of vector magnetic fields from Stokes profiles.}",
      journal = {\solphys},
         year = 1977,
        month = nov,
       volume = {55},
       number = {1},
        pages = {47-61},
          doi = {10.1007/BF00150873},
       adsurl = {https://ui.adsabs.harvard.edu/abs/1977SoPh...55...47A}
}

@ARTICLE{2007MmSAI..78..148L,
       author = {{Lites}, B. and {Casini}, R. and {Garcia}, J. and {Socas-Navarro}, H.},
        title = "{A suite of community tools for spectro-polarimetric analysis .}",
      journal = {\memsai},
         year = 2007,
        month = jan,
       volume = {78},
        pages = {148},
       adsurl = {https://ui.adsabs.harvard.edu/abs/2007MmSAI..78..148L}
}

@ARTICLE{2007A&A...462.1137O,
       author = {{Orozco Su{\'a}rez}, D. and {Del Toro Iniesta}, J.~C.},
        title = "{The usefulness of analytic response functions}",
      journal = {\aap},
         year = 2007,
        month = feb,
       volume = {462},
       number = {3},
        pages = {1137-1145},
          doi = {10.1051/0004-6361:20066201},
archivePrefix = {arXiv},
       eprint = {1211.1502},
 primaryClass = {astro-ph.SR},
       adsurl = {https://ui.adsabs.harvard.edu/abs/2007A&A...462.1137O}
}

@ARTICLE{2011SoPh..273..267B,
       author = {{Borrero}, J.~M. and {Tomczyk}, S. and {Kubo}, M. and {Socas-Navarro}, H. and {Schou}, J. and {Couvidat}, S. and {Bogart}, R.},
        title = "{VFISV: Very Fast Inversion of the Stokes Vector for the Helioseismic and Magnetic Imager}",
      journal = {\solphys},
         year = 2011,
        month = oct,
       volume = {273},
       number = {1},
        pages = {267-293},
          doi = {10.1007/s11207-010-9515-6},
archivePrefix = {arXiv},
       eprint = {0901.2702},
 primaryClass = {astro-ph.IM},
       adsurl = {https://ui.adsabs.harvard.edu/abs/2011SoPh..273..267B}
}

@ARTICLE{2019A&A...623A..74D,
       author = {{de la Cruz Rodr{\'\i}guez}, J. and {Leenaarts}, J. and {Danilovic}, S. and {Uitenbroek}, H.},
        title = "{STiC: A multiatom non-LTE PRD inversion code for full-Stokes solar observations}",
      journal = {\aap},
         year = 2019,
        month = mar,
       volume = {623},
          eid = {A74},
        pages = {A74},
          doi = {10.1051/0004-6361/201834464},
archivePrefix = {arXiv},
       eprint = {1810.08441},
 primaryClass = {astro-ph.SR},
       adsurl = {https://ui.adsabs.harvard.edu/abs/2019A&A...623A..74D}
}

@ARTICLE{2000A&A...358.1109F,
       author = {{Frutiger}, C. and {Solanki}, S.~K. and {Fligge}, M. and {Bruls}, J.~H.~M.~J.},
        title = "{Properties of the solar granulation obtained from the inversion of low spatial resolution spectra}",
      journal = {\aap},
         year = 2000,
        month = jun,
       volume = {358},
        pages = {1109-1121},
       adsurl = {https://ui.adsabs.harvard.edu/abs/2000A&A...358.1109F}
}

@ARTICLE{2018A&A...617A..24M,
       author = {{Mili{\'c}}, I. and {van Noort}, M.},
        title = "{Spectropolarimetric NLTE inversion code SNAPI}",
      journal = {\aap},
         year = 2018,
        month = sep,
       volume = {617},
          eid = {A24},
        pages = {A24},
          doi = {10.1051/0004-6361/201833382},
archivePrefix = {arXiv},
       eprint = {1806.08134},
 primaryClass = {astro-ph.SR},
       adsurl = {https://ui.adsabs.harvard.edu/abs/2018A&A...617A..24M}
}

@ARTICLE{2012A&A...548A...5V,
       author = {{van Noort}, M.},
        title = "{Spatially coupled inversion of spectro-polarimetric image data. I. Method and first results}",
      journal = {\aap},
         year = 2012,
        month = dec,
       volume = {548},
          eid = {A5},
        pages = {A5},
          doi = {10.1051/0004-6361/201220220},
archivePrefix = {arXiv},
       eprint = {1210.4636},
 primaryClass = {astro-ph.IM},
       adsurl = {https://ui.adsabs.harvard.edu/abs/2012A&A...548A...5V}
}

@ARTICLE{2019A&A...626A.102A,
       author = {{Asensio Ramos}, A. and {D{\'\i}az Baso}, C.~J.},
        title = "{Stokes inversion based on convolutional neural networks}",
      journal = {\aap},
         year = 2019,
        month = jun,
       volume = {626},
          eid = {A102},
        pages = {A102},
          doi = {10.1051/0004-6361/201935628},
archivePrefix = {arXiv},
       eprint = {1904.03714},
 primaryClass = {astro-ph.SR},
       adsurl = {https://ui.adsabs.harvard.edu/abs/2019A&A...626A.102A}
}

@ARTICLE{2022A&A...659A.165D,
       author = {{D{\'\i}az Baso}, C.~J. and {Asensio Ramos}, A. and {de la Cruz Rodr{\'\i}guez}, J.},
        title = "{Bayesian Stokes inversion with normalizing flows}",
      journal = {\aap},
         year = 2022,
        month = mar,
       volume = {659},
          eid = {A165},
        pages = {A165},
          doi = {10.1051/0004-6361/202142018},
archivePrefix = {arXiv},
       eprint = {2108.07089},
 primaryClass = {astro-ph.SR},
       adsurl = {https://ui.adsabs.harvard.edu/abs/2022A&A...659A.165D}
}

@ARTICLE{2022A&A...660A..37R,
       author = {{Ruiz Cobo}, B. and {Quintero Noda}, C. and {Gafeira}, R. and {Uitenbroek}, H. and {Orozco Su{\'a}rez}, D. and {P{\'a}ez Ma{\~n}{\'a}}, E.},
        title = "{DeSIRe: Departure coefficient aided Stokes Inversion based on Response functions}",
      journal = {\aap},
         year = 2022,
        month = apr,
       volume = {660},
          eid = {A37},
        pages = {A37},
          doi = {10.1051/0004-6361/202140877},
archivePrefix = {arXiv},
       eprint = {2202.02226},
 primaryClass = {astro-ph.SR},
       adsurl = {https://ui.adsabs.harvard.edu/abs/2022A&A...660A..37R}
}

@ARTICLE{2019A&A...629A..24P,
       author = {{Pastor Yabar}, A. and {Borrero}, J.~M. and {Ruiz Cobo}, B.},
        title = "{FIRTEZ-dz. A forward and inverse solver of the polarized radiative transfer equation under Zeeman regime in geometrical scale}",
      journal = {\aap},
         year = 2019,
        month = sep,
       volume = {629},
          eid = {A24},
        pages = {A24},
          doi = {10.1051/0004-6361/201935692},
archivePrefix = {arXiv},
       eprint = {1908.08075},
 primaryClass = {astro-ph.SR},
       adsurl = {https://ui.adsabs.harvard.edu/abs/2019A&A...629A..24P}
}

@ARTICLE{2019ApJ...873..128O,
       author = {{Osborne}, Christopher M.~J. and {Armstrong}, John A. and {Fletcher}, Lyndsay},
        title = "{RADYNVERSION: Learning to Invert a Solar Flare Atmosphere with Invertible Neural Networks}",
      journal = {\apj},
         year = 2019,
        month = mar,
       volume = {873},
       number = {2},
          eid = {128},
        pages = {128},
          doi = {10.3847/1538-4357/ab07b4},
archivePrefix = {arXiv},
       eprint = {1901.08626},
 primaryClass = {astro-ph.SR},
       adsurl = {https://ui.adsabs.harvard.edu/abs/2019ApJ...873..128O}
}

@ARTICLE{2018arXiv180804730A,
       author = {{Ardizzone}, Lynton and {Kruse}, Jakob and {Wirkert}, Sebastian and {Rahner}, Daniel and {Pellegrini}, Eric W. and {Klessen}, Ralf S. and {Maier-Hein}, Lena and {Rother}, Carsten and {K{\"o}the}, Ullrich},
        title = "{Analyzing Inverse Problems with Invertible Neural Networks}",
      journal = {arXiv e-prints},
         year = 2018,
        month = aug,
          eid = {arXiv:1808.04730},
        pages = {arXiv:1808.04730},
          doi = {10.48550/arXiv.1808.04730},
archivePrefix = {arXiv},
       eprint = {1808.04730},
 primaryClass = {cs.LG},
       adsurl = {https://ui.adsabs.harvard.edu/abs/2018arXiv180804730A}
}

@ARTICLE{2024ApJ...976..204Y,
       author = {{Yang}, Kai E. and {Tarr}, Lucas A. and {Rempel}, Matthias and {Dodds}, S. Curt and {Jaeggli}, Sarah A. and {Sadowski}, Peter and {Schad}, Thomas A. and {Cunnyngham}, Ian and {Liu}, Jiayi and {Glaser}, Yannik and {Sun}, Xudong},
        title = "{Spectropolarimetric Inversion in Four Dimensions with Deep Learning (SPIn4D). I. Overview, Magnetohydrodynamic Modeling, and Stokes Profile Synthesis}",
      journal = {\apj},
         year = 2024,
        month = dec,
       volume = {976},
       number = {2},
          eid = {204},
        pages = {204},
          doi = {10.3847/1538-4357/ad865b},
archivePrefix = {arXiv},
       eprint = {2407.20309},
 primaryClass = {astro-ph.SR},
       adsurl = {https://ui.adsabs.harvard.edu/abs/2024ApJ...976..204Y}
}

@ARTICLE{2024ApJ...977..101X,
       author = {{Xu}, Chong and {Wang}, JinLiang and {Li}, Hao and {Hu}, ZiYao and {Bai}, XianYong and {Lin}, JiaBen and {Liu}, Hui and {Jin}, ZhenYu and {Ji}, KaiFan},
        title = "{NNHMC: An Efficient Stokes Inversion Method Using a Neural Network (NN) Model Combined with the Hamiltonian Monte Carlo (HMC) Algorithm}",
      journal = {\apj},
         year = 2024,
        month = dec,
       volume = {977},
       number = {1},
          eid = {101},
        pages = {101},
          doi = {10.3847/1538-4357/ad8b2b},
       adsurl = {https://ui.adsabs.harvard.edu/abs/2024ApJ...977..101X}
}

@ARTICLE{2023SoPh..298...98M,
       author = {{Mistryukova}, Lukia and {Plotnikov}, Andrey and {Khizhik}, Aleksandr and {Knyazeva}, Irina and {Hushchyn}, Mikhail and {Derkach}, Denis},
        title = "{Stokes Inversion Techniques with Neural Networks: Analysis of Uncertainty in Parameter Estimation}",
      journal = {\solphys},
         year = 2023,
        month = aug,
       volume = {298},
       number = {8},
          eid = {98},
        pages = {98},
          doi = {10.1007/s11207-023-02189-4},
archivePrefix = {arXiv},
       eprint = {2210.14933},
 primaryClass = {astro-ph.SR},
       adsurl = {https://ui.adsabs.harvard.edu/abs/2023SoPh..298...98M}
}

@ARTICLE{2007A&A...476..959A,
       author = {{Asensio Ramos}, A. and {Mart{\'\i}nez Gonz{\'a}lez}, M.~J. and {Rubi{\~n}o-Mart{\'\i}n}, J.~A.},
        title = "{Bayesian inversion of Stokes profiles}",
      journal = {\aap},
         year = 2007,
        month = dec,
       volume = {476},
       number = {2},
        pages = {959-970},
          doi = {10.1051/0004-6361:20078107},
archivePrefix = {arXiv},
       eprint = {0709.0596},
 primaryClass = {astro-ph},
       adsurl = {https://ui.adsabs.harvard.edu/abs/2007A&A...476..959A}
}

@ARTICLE{2016LRSP...13....4D,
       author = {{del Toro Iniesta}, Jose Carlos and {Ruiz Cobo}, Basilio},
        title = "{Inversion of the radiative transfer equation for polarized light}",
      journal = {Living Reviews in Solar Physics},
         year = 2016,
        month = dec,
       volume = {13},
       number = {1},
          eid = {4},
        pages = {4},
          doi = {10.1007/s41116-016-0005-2},
archivePrefix = {arXiv},
       eprint = {1610.10039},
 primaryClass = {astro-ph.SR},
       adsurl = {https://ui.adsabs.harvard.edu/abs/2016LRSP...13....4D}
}

@ARTICLE{2017SSRv..210..109D,
       author = {{de la Cruz Rodr{\'\i}guez}, J. and {van Noort}, M.},
        title = "{Radiative Diagnostics in the Solar Photosphere and Chromosphere}",
      journal = {\ssr},
         year = 2017,
        month = sep,
       volume = {210},
       number = {1-4},
        pages = {109-143},
          doi = {10.1007/s11214-016-0294-8},
archivePrefix = {arXiv},
       eprint = {1609.08324},
 primaryClass = {astro-ph.SR},
       adsurl = {https://ui.adsabs.harvard.edu/abs/2017SSRv..210..109D}
}

@ARTICLE{2019A&A...631A.153D,
       author = {{de la Cruz Rodr{\'\i}guez}, J.},
        title = "{A method for global inversion of multi-resolution solar data}",
      journal = {\aap},
         year = 2019,
        month = nov,
       volume = {631},
          eid = {A153},
        pages = {A153},
          doi = {10.1051/0004-6361/201936635},
archivePrefix = {arXiv},
       eprint = {1909.02604},
 primaryClass = {astro-ph.SR},
       adsurl = {https://ui.adsabs.harvard.edu/abs/2019A&A...631A.153D}
}

@ARTICLE{2025A&A...703A..55A,
       author = {{Asensio Ramos}, A. and {de la Cruz Rodr{\'\i}guez}, J.},
        title = "{Neural translation for Stokes inversion and synthesis}",
      journal = {\aap},
         year = 2025,
        month = nov,
       volume = {703},
          eid = {A55},
        pages = {A55},
          doi = {10.1051/0004-6361/202556195},
archivePrefix = {arXiv},
       eprint = {2507.00594},
 primaryClass = {astro-ph.IM},
       adsurl = {https://ui.adsabs.harvard.edu/abs/2025A&A...703A..55A}
}

@article{Yang_2024,
doi = {10.3847/1538-4357/ad865b},
url = {https://doi.org/10.3847/1538-4357/ad865b},
year = {2024},
month = {nov},
publisher = {The American Astronomical Society},
volume = {976},
number = {2},
pages = {204},
author = {Yang, Kai E. and Tarr, Lucas A. and Rempel, Matthias and Dodds, S. Curt and Jaeggli, Sarah A. and Sadowski, Peter and Schad, Thomas A. and Cunnyngham, Ian and Liu, Jiayi and Glaser, Yannik and Sun, Xudong},
title = {Spectropolarimetric Inversion in Four Dimensions with Deep Learning (SPIn4D). I. Overview, Magnetohydrodynamic Modeling, and Stokes Profile Synthesis},
journal = {The Astrophysical Journal}
}

@ARTICLE{2012ApJ...750...62R,
       author = {{Rempel}, M.},
        title = "{Numerical Sunspot Models: Robustness of Photospheric Velocity and Magnetic Field Structure}",
      journal = {\apj},
         year = 2012,
        month = may,
       volume = {750},
       number = {1},
          eid = {62},
        pages = {62},
          doi = {10.1088/0004-637X/750/1/62},
archivePrefix = {arXiv},
       eprint = {1203.0534},
 primaryClass = {astro-ph.SR},
       adsurl = {https://ui.adsabs.harvard.edu/abs/2012ApJ...750...62R}
}

@article{unet15,  
  author = {Ronneberger, Olaf and Fischer, Philipp and Brox, Thomas},
  ee = {http://arxiv.org/abs/1505.04597},
  journal = {CoRR},  
  title = {U-Net: Convolutional Networks for Biomedical Image Segmentation.},
  url = {http://dblp.uni-trier.de/db/journals/corr/corr1505.html#RonnebergerFB15},
  volume = {abs/1505.04597},
  year = 2015
}

@article{lipman24,
  author = {Lipman, Yaron and Havasi, Marton and Holderrieth, Peter and Shaul, Neta and Le, Matt and Karrer, Brian and Chen, Ricky T. Q. and Lopez-Paz, David and Ben-Hamu, Heli and Gat, Itai},
  ee = {https://doi.org/10.48550/arXiv.2412.06264},
  journal = {CoRR},
  title = {Flow Matching Guide and Code.},
  url = {http://dblp.uni-trier.de/db/journals/corr/corr2412.html#abs-2412-06264},
  volume = {abs/2412.06264},
  year = 2024
}

@ARTICLE{1994SoPh..155..235M,
       author = {{Metcalf}, Thomas R.},
        title = "{Resolving the 180-degree ambiguity in vector magnetic field measurements: The `minimum' energy solution}",
      journal = {\solphys},
         year = 1994,
        month = dec,
       volume = {155},
       number = {2},
        pages = {235-242},
          doi = {10.1007/BF00680593},
       adsurl = {https://ui.adsabs.harvard.edu/abs/1994SoPh..155..235M}
}

@ARTICLE{2006SoPh..237..267M,
       author = {{Metcalf}, Thomas R. and {Leka}, K.~D. and {Barnes}, Graham and {Lites}, Bruce W. and {Georgoulis}, Manolis K. and {Pevtsov}, A.~A. and {Balasubramaniam}, K.~S. and {Gary}, G. Allen and {Jing}, Ju and {Li}, Jing and {Liu}, Y. and {Wang}, H.~N. and {Abramenko}, Valentyna and {Yurchyshyn}, Vasyl and {Moon}, Y.-J.},
        title = "{An Overview of Existing Algorithms for Resolving the 180$^{{\textdegree}}$ Ambiguity in Vector Magnetic Fields: Quantitative Tests with Synthetic Data}",
      journal = {\solphys},
         year = 2006,
        month = sep,
       volume = {237},
       number = {2},
        pages = {267-296},
          doi = {10.1007/s11207-006-0170-x},
       adsurl = {https://ui.adsabs.harvard.edu/abs/2006SoPh..237..267M}
}

@ARTICLE{2008ApJ...672.1237L,
       author = {{Lites}, B.~W. and {Kubo}, M. and {Socas-Navarro}, H. and {Berger}, T. and {Frank}, Z. and {Shine}, R. and {Tarbell}, T. and {Title}, A. and {Ichimoto}, K. and {Katsukawa}, Y. and {Tsuneta}, S. and {Suematsu}, Y. and {Shimizu}, T. and {Nagata}, S.},
        title = "{The Horizontal Magnetic Flux of the Quiet-Sun Internetwork as Observed with the Hinode Spectro-Polarimeter}",
      journal = {\apj},
         year = 2008,
        month = jan,
       volume = {672},
       number = {2},
        pages = {1237-1253},
          doi = {10.1086/522922},
       adsurl = {https://ui.adsabs.harvard.edu/abs/2008ApJ...672.1237L}
}

@ARTICLE{2023A&A...669A.122B,
       author = {{Borrero}, J.~M. and {Pastor Yabar}, A.},
        title = "{Combining magneto-hydrostatic constraints with Stokes profiles inversions. III. Uncertainty in the inference of electric currents}",
      journal = {\aap},
         year = 2023,
        month = jan,
       volume = {669},
          eid = {A122},
        pages = {A122},
          doi = {10.1051/0004-6361/202244716},
archivePrefix = {arXiv},
       eprint = {2211.07593},
 primaryClass = {astro-ph.SR},
       adsurl = {https://ui.adsabs.harvard.edu/abs/2023A&A...669A.122B}
}

@ARTICLE{1999Ap&SS.264...77P,
       author = {{Priest}, E.~R.},
        title = "{Heating the Solar Corona by Magnetic Reconnection}",
      journal = {\apss},
         year = 1999,
        month = jan,
       volume = {264},
        pages = {77-100},
          doi = {10.1023/A:1002440524834},
       adsurl = {https://ui.adsabs.harvard.edu/abs/1999Ap&SS.264...77P}
}

@ARTICLE{2012ApJ...747...87K,
       author = {{Khomenko}, E. and {Collados}, M.},
        title = "{Heating of the Magnetized Solar Chromosphere by Partial Ionization Effects}",
      journal = {\apj},
         year = 2012,
        month = mar,
       volume = {747},
       number = {2},
          eid = {87},
        pages = {87},
          doi = {10.1088/0004-637X/747/2/87},
archivePrefix = {arXiv},
       eprint = {1112.3374},
 primaryClass = {astro-ph.SR},
       adsurl = {https://ui.adsabs.harvard.edu/abs/2012ApJ...747...87K}
}

@ARTICLE{2021A&A...656L..20P,
       author = {{Pastor Yabar}, A. and {Borrero}, J.~M. and {Quintero Noda}, C. and {Ruiz Cobo}, B.},
        title = "{Inference of electric currents in the solar photosphere}",
      journal = {\aap},
         year = 2021,
        month = dec,
       volume = {656},
          eid = {L20},
        pages = {L20},
          doi = {10.1051/0004-6361/202142149},
archivePrefix = {arXiv},
       eprint = {2112.04356},
 primaryClass = {astro-ph.SR},
       adsurl = {https://ui.adsabs.harvard.edu/abs/2021A&A...656L..20P}
}

@ARTICLE{2021A&A...647A.190B,
       author = {{Borrero}, J.~M. and {Pastor Yabar}, A. and {Ruiz Cobo}, B.},
        title = "{Combining magneto-hydrostatic constraints with Stokes profiles inversions. II. Application to Hinode/SP observations}",
      journal = {\aap},
         year = 2021,
        month = mar,
       volume = {647},
          eid = {A190},
        pages = {A190},
          doi = {10.1051/0004-6361/202039927},
archivePrefix = {arXiv},
       eprint = {2101.04394},
 primaryClass = {astro-ph.SR},
       adsurl = {https://ui.adsabs.harvard.edu/abs/2021A&A...647A.190B}
}

@ARTICLE{2009ApJ...700.1391M,
       author = {{Mart{\'\i}nez Gonz{\'a}lez}, M.~J. and {Bellot Rubio}, L.~R.},
        title = "{Emergence of Small-scale Magnetic Loops Through the Quiet Solar Atmosphere}",
      journal = {\apj},
         year = 2009,
        month = aug,
       volume = {700},
       number = {2},
        pages = {1391-1403},
          doi = {10.1088/0004-637X/700/2/1391},
archivePrefix = {arXiv},
       eprint = {0905.2691},
 primaryClass = {astro-ph.SR},
       adsurl = {https://ui.adsabs.harvard.edu/abs/2009ApJ...700.1391M}
}

@ARTICLE{2022A&A...666A..21Q,
       author = {{Quintero Noda}, C. and {Schlichenmaier}, R. and {Bellot Rubio}, L.~R. and {L{\"o}fdahl}, M.~G. and {Khomenko}, E. and {Jur{\v{c}}{\'a}k}, J. and {Leenaarts}, J. and {Kuckein}, C. and {Gonz{\'a}lez Manrique}, S.~J. and {Gun{\'a}r}, S. and {Nelson}, C.~J. and {de la Cruz Rodr{\'\i}guez}, J. and {Tziotziou}, K. and {Tsiropoula}, G. and {Aulanier}, G. and {Aboudarham}, J. and {Allegri}, D. and {Alsina Ballester}, E. and {Amans}, J.~P. and {Asensio Ramos}, A. and {Bail{\'e}n}, F.~J. and {Balaguer}, M. and {Baldini}, V. and {Balthasar}, H. and {Barata}, T. and {Barczynski}, K. and {Barreto Cabrera}, M. and {Baur}, A. and {B{\'e}chet}, C. and {Beck}, C. and {Bel{\'\i}o-As{\'\i}n}, M. and {Bello-Gonz{\'a}lez}, N. and {Belluzzi}, L. and {Bentley}, R.~D. and {Berdyugina}, S.~V. and {Berghmans}, D. and {Berlicki}, A. and {Berrilli}, F. and {Berkefeld}, T. and {Bettonvil}, F. and {Bianda}, M. and {Bienes P{\'e}rez}, J. and {Bonaque-Gonz{\'a}lez}, S. and {Braj{\v{s}}a}, R. and {Bommier}, V. and {Bourdin}, P.-A. and {Burgos Mart{\'\i}n}, J. and {Calchetti}, D. and {Calcines}, A. and {Calvo Tovar}, J. and {Campbell}, R.~J. and {Carballo-Mart{\'\i}n}, Y. and {Carbone}, V. and {Carlin}, E.~S. and {Carlsson}, M. and {Castro L{\'o}pez}, J. and {Cavaller}, L. and {Cavallini}, F. and {Cauzzi}, G. and {Cecconi}, M. and {Chulani}, H.~M. and {Cirami}, R. and {Consolini}, G. and {Coretti}, I. and {Cosentino}, R. and {C{\'o}zar-Castellano}, J. and {Dalmasse}, K. and {Danilovic}, S. and {De Juan Ovelar}, M. and {Del Moro}, D. and {del Pino Alem{\'a}n}, T. and {del Toro Iniesta}, J.~C. and {Denker}, C. and {Dhara}, S.~K. and {Di Marcantonio}, P. and {D{\'\i}az Baso}, C.~J. and {Diercke}, A. and {Dineva}, E. and {D{\'\i}az-Garc{\'\i}a}, J.~J. and {Doerr}, H.-P. and {Doyle}, G. and {Erdelyi}, R. and {Ermolli}, I. and {Escobar Rodr{\'\i}guez}, A. and {Esteban Pozuelo}, S. and {Faurobert}, M. and {Felipe}, T. and {Feller}, A. and {Feijoo Amoedo}, N. and {Femen{\'\i}a Castell{\'a}}, B. and {Fernandes}, J. and {Ferro Rodr{\'\i}guez}, I. and {Figueroa}, I. and {Fletcher}, L. and {Franco Ordovas}, A. and {Gafeira}, R. and {Gardenghi}, R. and {Gelly}, B. and {Giorgi}, F. and {Gisler}, D. and {Giovannelli}, L. and {Gonz{\'a}lez}, F. and {Gonz{\'a}lez}, J.~B. and {Gonz{\'a}lez-Cava}, J.~M. and {Gonz{\'a}lez Garc{\'\i}a}, M. and {G{\"o}m{\"o}ry}, P. and {Gracia}, F. and {Grauf}, B. and {Greco}, V. and {Grivel}, C. and {Guerreiro}, N. and {Guglielmino}, S.~L. and {Hammerschlag}, R. and {Hanslmeier}, A. and {Hansteen}, V. and {Heinzel}, P. and {Hern{\'a}ndez-Delgado}, A. and {Hern{\'a}ndez Su{\'a}rez}, E. and {Hidalgo}, S.~L. and {Hill}, F. and {Hizberger}, J. and {Hofmeister}, S. and {J{\"a}gers}, A. and {Janett}, G. and {Jarolim}, R. and {Jess}, D. and {Jim{\'e}nez Mej{\'\i}as}, D. and {Jolissaint}, L. and {Kamlah}, R. and {Kapit{\'a}n}, J. and {Ka{\v{s}}parov{\'a}}, J. and {Keller}, C.~U. and {Kentischer}, T. and {Kiselman}, D. and {Kleint}, L. and {Klvana}, M. and {Kontogiannis}, I. and {Krishnappa}, N. and {Ku{\v{c}}era}, A. and {Labrosse}, N. and {Lagg}, A. and {Landi Degl'Innocenti}, E. and {Langlois}, M. and {Lafon}, M. and {Laforgue}, D. and {Le Men}, C. and {Lepori}, B. and {Lepreti}, F. and {Lindberg}, B. and {Lilje}, P.~B. and {L{\'o}pez Ariste}, A. and {L{\'o}pez Fern{\'a}ndez}, V.~A. and {L{\'o}pez Jim{\'e}nez}, A.~C. and {L{\'o}pez L{\'o}pez}, R. and {Manso Sainz}, R. and {Marassi}, A. and {Marco de la Rosa}, J. and {Marino}, J. and {Marrero}, J. and {Mart{\'\i}n}, A. and {Mart{\'\i}n G{\'a}lvez}, A. and {Mart{\'\i}n Hernando}, Y. and {Masciadri}, E. and {Mart{\'\i}nez Gonz{\'a}lez}, M. and {Matta-G{\'o}mez}, A. and {Mato}, A. and {Mathioudakis}, M. and {Matthews}, S. and {Mein}, P. and {Merlos Garc{\'\i}a}, F. and {Moity}, J. and {Montilla}, I. and {Molinaro}, M. and {Molodij}, G. and {Montoya}, L.~M. and {Munari}, M. and {Murabito}, M. and {N{\'u}{\~n}ez Cagigal}, M. and {Oliviero}, M. and {Orozco Su{\'a}rez}, D. and {Ortiz}, A. and {Padilla-Hern{\'a}ndez}, C. and {Pa{\'e}z Ma{\~n}{\'a}}, E. and {Paletou}, F. and {Pancorbo}, J. and {Pastor Ca{\~n}edo}, A. and {Pastor Yabar}, A. and {Peat}, A.~W. and {Pedichini}, F. and {Peixinho}, N. and {Pe{\~n}ate}, J. and {P{\'e}rez de Taoro}, A. and {Peter}, H. and {Petrovay}, K. and {Piazzesi}, R. and {Pietropaolo}, E. and {Pleier}, O. and {Poedts}, S. and {P{\"o}tzi}, W. and {Podladchikova}, T.},
        title = "{The European Solar Telescope}",
      journal = {\aap},
         year = 2022,
        month = oct,
       volume = {666},
          eid = {A21},
        pages = {A21},
          doi = {10.1051/0004-6361/202243867},
archivePrefix = {arXiv},
       eprint = {2207.10905},
 primaryClass = {astro-ph.SR},
       adsurl = {https://ui.adsabs.harvard.edu/abs/2022A&A...666A..21Q}
}

@inproceedings{rombach2022high,
  author = {Rombach, Robin and Blattmann, Andreas and Lorenz, Dominik and Esser, Patrick and Ommer, Bj{\"o}rn},
  booktitle = {Proceedings of the IEEE/CVF Conference on Computer Vision and Pattern Recognition},
  pages = {10684--10695},
  title = {High-resolution image synthesis with latent diffusion models},
  year = 2022
}

@ARTICLE{2017A&A...604A..11A,
       author = {{Asensio Ramos}, A. and {Requerey}, I.~S. and {Vitas}, N.},
        title = "{DeepVel: Deep learning for the estimation of horizontal velocities at the solar surface}",
      journal = {\aap},
         year = 2017,
        month = jul,
       volume = {604},
          eid = {A11},
        pages = {A11},
          doi = {10.1051/0004-6361/201730783},
archivePrefix = {arXiv},
       eprint = {1703.05128},
 primaryClass = {astro-ph.SR},
       adsurl = {https://ui.adsabs.harvard.edu/abs/2017A&A...604A..11A}
}

@ARTICLE{2010ApJ...721L..58P,
       author = {{Puschmann}, K.~G. and {Ruiz Cobo}, B. and {Mart{\'\i}nez Pillet}, V.},
        title = "{The Electrical Current Density Vector in the Inner Penumbra of a Sunspot}",
      journal = {\apjl},
         year = 2010,
        month = sep,
       volume = {721},
       number = {1},
        pages = {L58-L61},
          doi = {10.1088/2041-8205/721/1/L58},
archivePrefix = {arXiv},
       eprint = {1008.2131},
 primaryClass = {astro-ph.SR},
       adsurl = {https://ui.adsabs.harvard.edu/abs/2010ApJ...721L..58P}
}

@ARTICLE{2010ApJ...720.1417P,
       author = {{Puschmann}, K.~G. and {Ruiz Cobo}, B. and {Mart{\'\i}nez Pillet}, V.},
        title = "{A Geometrical Height Scale for Sunspot Penumbrae}",
      journal = {\apj},
         year = 2010,
        month = sep,
       volume = {720},
       number = {2},
        pages = {1417-1431},
          doi = {10.1088/0004-637X/720/2/1417},
archivePrefix = {arXiv},
       eprint = {1007.2779},
 primaryClass = {astro-ph.SR},
       adsurl = {https://ui.adsabs.harvard.edu/abs/2010ApJ...720.1417P}
}

@ARTICLE{2022ApJ...933..145L,
       author = {{Li}, H. and {del Pino Alem{\'a}n}, T. and {Trujillo Bueno}, J. and {Casini}, R.},
        title = "{TIC: A Stokes Inversion Code for Scattering Polarization with Partial Frequency Redistribution and Arbitrary Magnetic Fields}",
      journal = {\apj},
         year = 2022,
        month = jul,
       volume = {933},
       number = {2},
          eid = {145},
        pages = {145},
          doi = {10.3847/1538-4357/ac745c},
archivePrefix = {arXiv},
       eprint = {2205.15666},
 primaryClass = {astro-ph.SR},
       adsurl = {https://ui.adsabs.harvard.edu/abs/2022ApJ...933..145L}
}

@ARTICLE{2025ApJ...995..146Y,
       author = {{Yang}, Kai E. and {Sun}, Xudong and {Tarr}, Lucas A. and {Liu}, Jiayi and {Sadowski}, Peter and {Dodds}, S. Curt and {Rempel}, Matthias and {Jaeggli}, Sarah A. and {Schad}, Thomas A. and {Cunnyngham}, Ian and {Glaser}, Yannik and {Wolniewicz}, Linnea},
        title = "{Spectropolarimetric Inversion in Four Dimensions with Deep Learning (SPIn4D). II. A Physics-informed Machine Learning Method for 3D Solar Photosphere Reconstruction}",
      journal = {\apj},
         year = 2025,
        month = dec,
       volume = {995},
       number = {2},
          eid = {146},
        pages = {146},
          doi = {10.3847/1538-4357/ae12ef},
archivePrefix = {arXiv},
       eprint = {2510.09967},
 primaryClass = {astro-ph.SR},
       adsurl = {https://ui.adsabs.harvard.edu/abs/2025ApJ...995..146Y}
}

@unpublished{Nick2026,
  author = {{Glaser}, Yannik and {Sadowski}, Peter and {Yang}, Kai and {Sun}, Xudong and {Tarr}, Lucas A. and {Liu}, Jiayi and {Dodds}, S. Curt and {Rempel}, Matthias and {Jaeggli}, Sarah A. and {Schad}, Thomas A. and {Cunnyngham}, Ian and {Wolniewicz}, Linnea},
  title = {{Spectropolarimetric Inversion in Four Dimensions with Deep Learning ({SPIn4D}). III.}},
  note = {in preparation},
  year = {2026}
}

\begin{appendix}
\section{Disambiguation of the transverse magnetic field}
The disambiguation of the transverse components of the magnetic field is a crucial step in 
the inference of the magnetic field vector, since the Zeeman effect only provides information about the azimuth of 
the magnetic field modulo 180$^\circ$. We have developed a smooth version of the minimum energy method of
\cite{1994SoPh..155..235M}.
We make use of the inferred $B_{p1}$ and $B_{p2}$ to compute the still ambiguous solution of $B_x$ and $B_y$ whose azimuth
lines in the range $[0,\pi]$. The azimuth is computed as follows:
\begin{equation}
\phi' = \frac{1}{2} \arctan \frac{B_{p2}}{B_{p1}} \mod \pi,
\end{equation}
which can be used to calculate the transverse component of the magnetic field as 
$B_x=S B_t \cos \phi'$ and $B_y=S B_t \sin \phi'$, where $S=\pm 1$ is the switch that determines the solution of the ambiguity. 
Strictly speaking, the disambiguation process is a 
combinatorial problem, since $S$ at each pixel has two possible solutions, resulting in the two
options ($B_x, B_y$) and ($-B_x, -B_y$). However, the problem can be relaxed to a continuous optimization, which
can be solved efficiently with gradient-based methods. This is achieved
by defining the per-pixel variable $\theta$ and using $S=\tanh \theta$ as a soft switch between the two 
possible solutions. The variable $\theta$ is then optimized by minimizing the 
following criterion:
\begin{equation}
E = \sum_{i} \left( \left| \nabla \cdot \mathbf{B} \right| + \lambda \left| J \right| \right)^2,
\end{equation}
where the summation of the criterion is carried out over all pixels in the FoV. The parameter $\lambda$ is a hyperparameter 
that controls the relative importance of the two terms in the criterion, which we choose $\lambda=1$.
We again use the Adam optimizer with a learning rate of $10^{-2}$ to optimize the variable $\theta$ for 5000 iterations.
After several random initializations, we find that the final solution is very robust and almost insensitive 
to the initial conditions.

During the optimization, the magnetic field divergence and the cartesian components of the 
electric current density $\mathbf{J}$, given in A\,m$^{-2}$, are calculated using Ampere's law as:
\begin{align}
\nabla \cdot \mathbf{B} &= \frac{\partial B_x}{\partial x} + \frac{\partial B_y}{\partial y} + \frac{\partial B_z}{\partial z} \\
J_x &= \mu_0^{-1} \left( \frac{\partial B_z}{\partial y} - \frac{\partial B_y}{\partial z} \right) \nonumber \\
J_y &= \mu_0^{-1} \left( \frac{\partial B_x}{\partial z} - \frac{\partial B_z}{\partial x} \right) \nonumber \\
J_z &= \mu_0^{-1} \left( \frac{\partial B_y}{\partial x} - \frac{\partial B_x}{\partial y} \right) \\
\end{align}
where $\mu_0=4\pi \times 10^{-7}$ kg\,m\,s$^{-2}$\,A$^{-2}$ is the magnetic permeability of free space
and the magnetic fields are measured in T. The horizontal spatial derivatives are calculated using 
finite differences. The vertical derivatives are calculated using the inferred atmospheric parameters at 
different heights in the atmosphere. For this purpose, we fix 100 km as the target height for the disambiguation
and then compute the vertical derivatives by interpolation at 90 and 100 km. The code is part of the
\code package and is publicly available at \urlcode.
\end{appendix}

\end{document}